\documentclass[aps,prd,preprint,nofootinbib]{revtex4}
\usepackage{mathrsfs}
\usepackage{amsfonts}
\usepackage{amsmath}
\usepackage{array}
\usepackage{epsfig}
\usepackage{graphicx}
\newcommand{\bqa}{\begin{eqnarray}}
\newcommand{\eqa}{\end{eqnarray}}

\begin{document}
\title{NLO QCD Corrections to $B_c$-to-Charmonium Form Factors \\[9mm]}

\author{Cong-Feng Qiao$^{1}$ and Peng Sun$^{1,2}$, and Feng Yuan$^{3}$}
\affiliation{$^{1}$College of Physical
Sciences, Graduate University of Chinese Academy of Sciences \\
YuQuan Road 19A, Beijing 100049, China} \affiliation{$^{2}$Center
for High-Energy Physics, Peking University, Beijing 100871, China}
\affiliation{$^{3}$Nuclear Science Division, LBNL, Berkeley, CA
94720, USA}
\author{~\vspace{0.9cm}}

\begin{abstract}
\vspace{3mm} The $B_c(^1S_0)$ meson to S-wave Charmonia transition
form factors in large recoil region are calculated in
next-to-leading order(NLO) accuracy of Quantum Chromodynamics(QCD).
Our results indicate that the higher order corrections to these form
factors are remarkable, and hence are important to the
phenomenological study of the corresponding processes. For the
convenience of comparison and use, the relevant expressions in
asymptotic form in the limit of $m_c\rightarrow0$ are presented.

\vspace {7mm} \noindent {\bf PACS number(s):} 12.38.Bx, 12.38.St,
12.39.Hg, 14.40.Nd.

\end{abstract}
 \maketitle
\section{Introduction}
The study of $B_c$ meson is of special interest, since it is the
only heavy meson composed of two heavy quarks with different
flavors. The $B_c$ exclusive decays provide an important
non-relativistic system in the investigation of weak interaction,
hadronic properties of heavy mesons and even new physics. Up to now
there are only two decay modes of $B_c$ meson being observed in
experiment at the Fermilab Tevatron, i.e. $B_c(^1S_0) \rightarrow
J/\psi\pi$ and $B_c(^1S_0) \rightarrow J/\psi e^+\nu_e$ \cite{exp1}.
Theoretically, many of works were carried out in different
frameworks, see for instance recent works \cite{bib1,bib2} and
references therein. In analyzing the $B_c$ decay processes, there
are several different scales should be taken into account: the hard
scale set by the heavy quark masses $m_Q$, the soft scale set by
$m_Qv$ where $v<1$ is the relative velocity of heavy quarks within
the $B_c$ meson, and the ultrasoft scale set by $m_Qv^2$. The hard
part supplies the short-distance contribution and can be calculated
perturbatively in strong interaction, while the soft and ultrasoft
parts belong to the long-distance contribution and have to be
evaluated via some non-perturbative methods or fitted by
experimental data.

In the study of $B$-meson decays, the factorization \cite{Fac1,Fac2}
is crucial to disentangle the short-distance sector from the
long-distance sector, where the former can be treated by
perturbertive QCD(pQCD), while the later can be characterized by
some universal hadronic parameters. Because $B$ to light hadron
exclusive decays are mediated by weak interaction, it is convenient
to use an effective weak Hamiltonian to describe the interaction,
which has the following structure:
\begin{eqnarray}
\mathcal
{H}_{eff}=\frac{G_F}{\sqrt{2}}\sum_iV^{i}_{CKM}C_i(\mu)Q_i\; .
\end{eqnarray}
Here $G_F$ is the Fermi constant and $Q_i$ are local operators,
$C_i$ are short-distance coefficients \cite{wilson1,wilson2} and
$V^{i}_{CKM}$ is CKM matrix element \cite{CKM1,CKM2}. In naive
factorization approach, the $B$-meson exclusive two-body decays can
be formulated as
\begin{eqnarray}
\langle M_1 M_2|Q_i|B\rangle\sim\langle M_1|\overline{\psi} \Gamma
b|B\rangle\langle M_2|\overline{\psi}\Gamma\psi|0\rangle\; ,
\end{eqnarray}
where the matrix element $\langle M_1|\overline{\psi} \Gamma
b|B\rangle$ stands for the transition form factor at large recoil,
and $\langle M_2|\overline{\psi}\Gamma\psi|0\rangle$ corresponds to
the $M_2$ decay constant. If we consider all the partons on
light-cone, the matrix element $\langle M_1|\overline{\psi} \Gamma
b|B\rangle$ can not be factorized further due to the divergence at
the end point arising from vanishing energy of the partons on the
light-cone. However, if we extend to the non-relativistic situation,
the nonperturbative effect can be factorized to Coulomb potentials
of initial or final bound states. For instance, in the process
$B_c\rightarrow J/\psi(\eta_c) + \pi$, we can describe the dynamics
of bound states $B_c$ and $J/\psi(\eta_c)$ by non-relativistic
QCD(NRQCD) \cite{NRQCD}, since the masses of bottom and charm quarks
are much bigger than $\Lambda_{QCD}$. And then, the matrix element
relevant to the form factor at large recoil can be factorized as
\cite{Bc1,Bc2}:
\begin{eqnarray}
\langle J/\psi(\eta_c)|\overline{c} \Gamma_i
b|B_c\rangle\simeq\psi_{B_c}(0)\psi_{J/\psi(\eta_c)}(0)T_i\; .
\end{eqnarray}
Here, the nonperturbative parameters $\psi_{\bar{B}_c}(0)$ and
$\psi_{J/\psi(\eta_c)}(0)$ are the Schr\"{o}dinger wave functions at
the origin for $b\bar{c}$ and $c\bar{c}$ systems, respectively.
$T_i$ is a hard scattering kernel which can be calculated
perturbatively.

As the LHC will soon be in the position to explore many $B_c$ decay
channels - among these several are in semileptonic and nonleptonic
charmonium decay modes - a dedicated study of $B_c$-to-charmonium
form factors is meaningful. In this work, we will explicitly
calculate the matrix elements $\langle J/\psi|\overline{c}
\Gamma_{V(A)} b|B_c\rangle$ and $\langle\eta_c|\overline{c}
\Gamma_{V} b|B_c\rangle$ in pQCD approach at the next-to-leading
order in non-relativistic limit of the initial and final bound
states. The paper is organized as follows: in section II, we
represent matrix element at the Born level; in section III, we
calculate the matrix elements in the NLO accuracy in pQCD; in
section IV, we compare result from pQCD with the wave-function
overlap contribution qualitatively; in the last section a brief
summary and conclusions are given.

\section{The Form Factors at Born Level}
\begin{figure}[h,m,u]
\centering
\includegraphics[width=10cm,height=3cm]{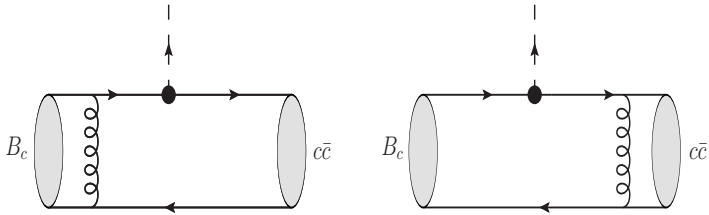}%
\caption{\small The leading order Feynman diagrams } \label{graph1}
\end{figure}
The investigation of process $B_c$ decays to S-wave charmonia
($J/\psi$ or $\eta_c$) with a light meson or lepton pair plays an
important role in the study of $B_c$ property, where the nature of
the transition form factor stands as a central issue. In this work,
we focus on the study of two and four independent form factors in
$B_c(^1S_0)$ to $\eta_c$ and $J/\psi$ transitions respectively,
which are normally defined as:
\begin{eqnarray}
\langle\eta_c(P')|\overline{c}\gamma^\mu
b|B_c(P)\rangle&=& f_+(P'+P)^\mu+f_-(P-P')^\mu\; ,\\
\langle J/\psi(P',\epsilon^*)|\overline{c}\gamma^\mu
b|B_c(P)\rangle&=&ig\epsilon^{\mu\nu\sigma\rho}\epsilon^*_\nu
P_\sigma P'_\rho\; ,\\
\langle J/\psi(P',\epsilon^*)|\overline{c}\gamma^\mu\gamma^5
b|B_c(P)\rangle& = & a_0\epsilon^{*\mu}+a_+\epsilon^*\cdot
PP'^\mu+a_-\epsilon^*\cdot PP^\mu \; .
\end{eqnarray}
At the leading order in $\alpha_{s}$, there are two independent
Feynman Diagrams for $\langle J/\psi|\overline{c} \Gamma_{V(A)}
b|B_c\rangle$ and $\langle\eta_c|\overline{c} \Gamma_{V(A)}
b|B_c\rangle$, as schemetically shown in Figure 1. In
non-relativistic limit, the momenta of constituent bottom and charm
quarks are $p_b=\xi P$ and $p_{\bar{c}}=(1-\xi)P$ with
$\xi=\frac{m_b}{m_c+m_b}$ for $B_c$ meson, and
$p_{\bar{c}}=p_c=P'/2$ for $J/\psi$ or $\eta_c$ meson. Here, $P$ and
$P'$ signify the momenta of initial $B_c$ and final charmonia.

After taking the above mentioned procedures, it is straightforward
to calculate those concerned form factors at the tree level. They
read
\bqa
f_+^{LO}=\frac{8\sqrt{2}\pi\alpha_s\psi(0)_{B_c}
\psi(0)_{\eta_c}C_AC_F(\sqrt{m_b+m_c})
(3m_b^2+2m_cm_b+3m_c^2-q^2)}{N_cm_c^{3/2}
(m_b^2+m_c^2-2m_bm_c-q^2)^2}\;
, \eqa
\bqa
f_-^{LO}=-\frac{16\sqrt{2}\pi\alpha_s\psi(0)_{B_c}
\psi(0)_{\eta_c}C_AC_F(m_b+m_c)^{3/2}
(m_b-m_c)}{N_cm_c^{3/2}(m_b^2+m_c^2-2m_cm_b-q^2)^2}\; ,
 \eqa
\bqa
g^{LO}=-\frac{32\sqrt{2}\pi\alpha_s
\psi(0)_{B_c}\psi(0)_{J/\psi}C_AC_F
(m_b+m_c)^{3/2}}{N_cm_c^{3/2}(m_b^2+m_c^2-2m_cm_b-q^2)^2}\; ,
 \eqa
\bqa
a_0^{LO}=\frac{16\sqrt{2}\pi\alpha_s\psi(0)_{B_c}
\psi(0)_{J/\psi}C_AC_F\sqrt{m_b+m_c}
(m_b^3+6m_cm_b^2+5m_c^2m_b-q^2m_b+4m_c^3-2m_cq^2)}
{N_cm_c^{3/2}(m_b^2+m_c^2-2m_cm_b-q^2)^2}\; ,\nonumber\\
 \eqa
\bqa
a_+^{LO}=-\frac{32\sqrt{2}\pi\alpha_s
\psi(0)_{B_c}\psi(0)_{J/\psi}C_AC_F
(m_b+m_c)^{3/2}}{N_cm_c^{3/2}(m_b^2+m_c^2-2m_cm_b-q^2)^2}\; ,
 \eqa
\bqa
a_-^{LO}=\frac{32\sqrt{2}\pi\alpha_s
\psi(0)_{B_c}\psi(0)_{J/\psi}C_AC_F
\sqrt{m_b+m_c}}{N_c\sqrt{m_c}(m_b^2+m_c^2-2m_cm_b-q^2)^2}\; .
 \eqa
Here, the momentum transfer $q = P-P'$, and the invariant mass $q^2
\rightarrow 0$, i.e. the finial charmonium owns the maximal
momentum, denotes the maximum recoil point.

\section{The Next-to-Leading Order Corrections}
\begin{figure}[h,m,u]
\centering
\includegraphics[width=10cm,height=7cm]{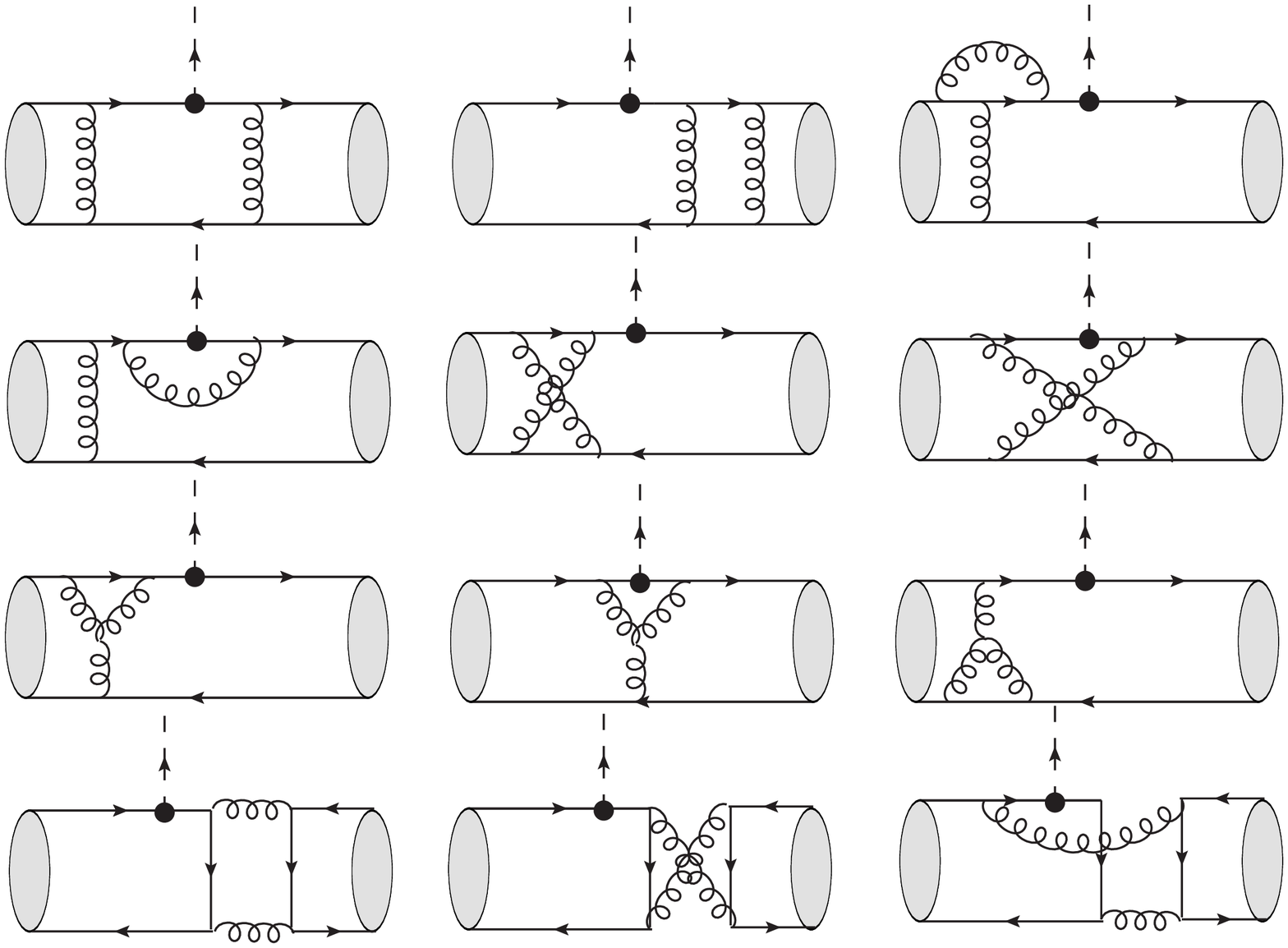}%
\caption{\small The typical Feynman diagrams at one-loop level.}
\label{graph2}
\end{figure}

\begin{figure}[htb]
\centering\vspace*{-1cm}
\includegraphics[width=0.5\textwidth]{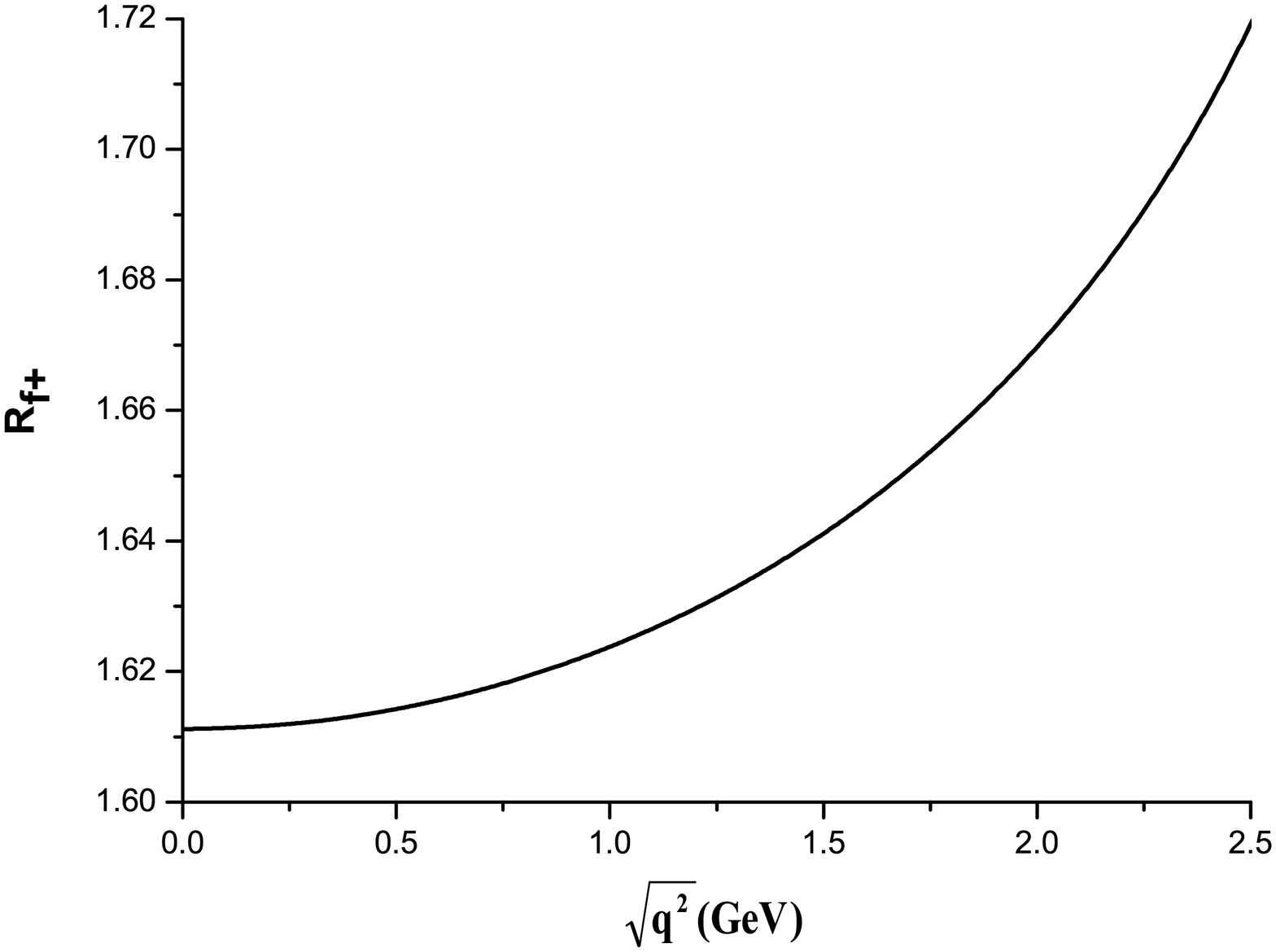}%
\includegraphics[width=0.5\textwidth]{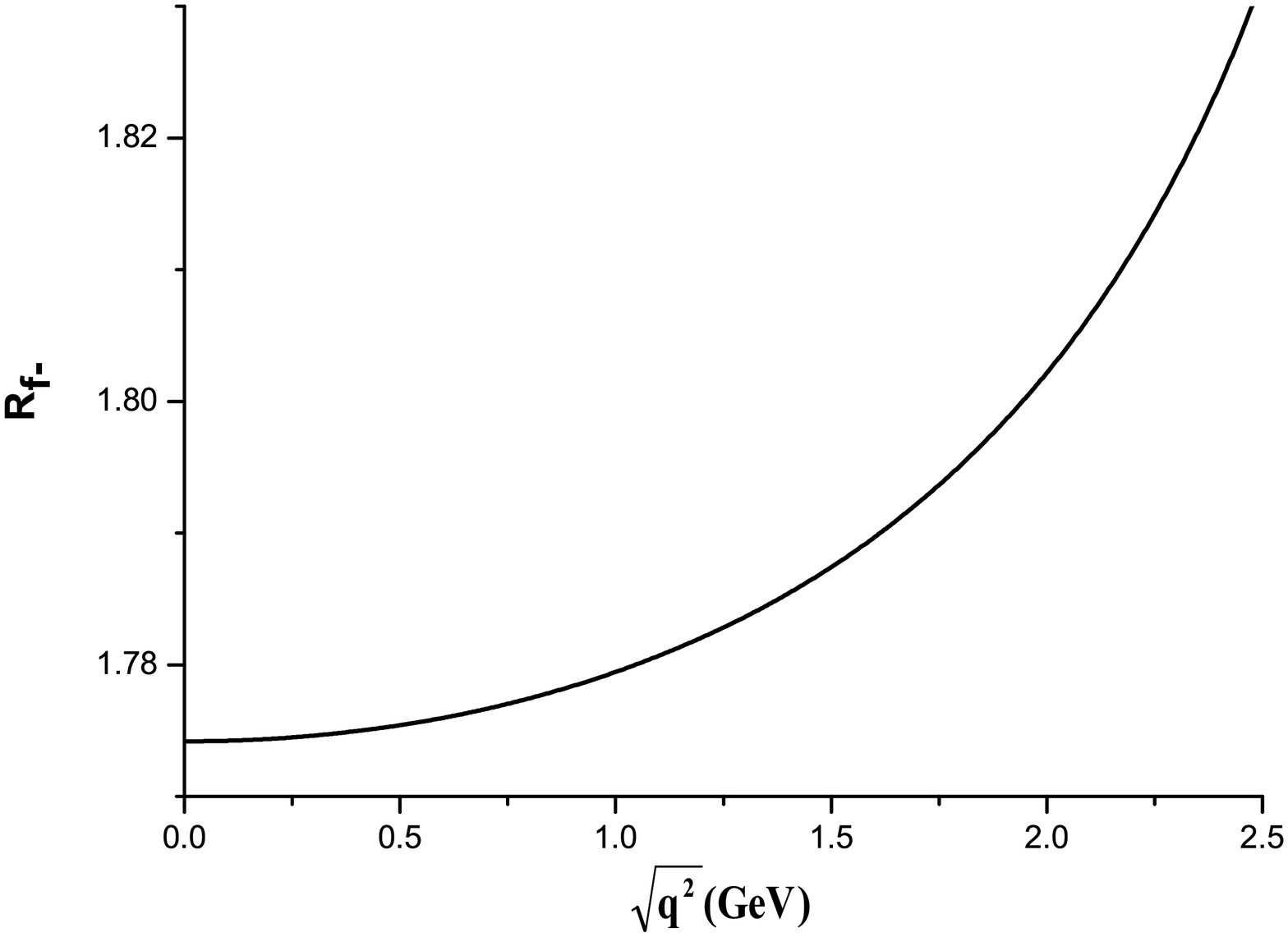}
\includegraphics[width=0.5\textwidth]{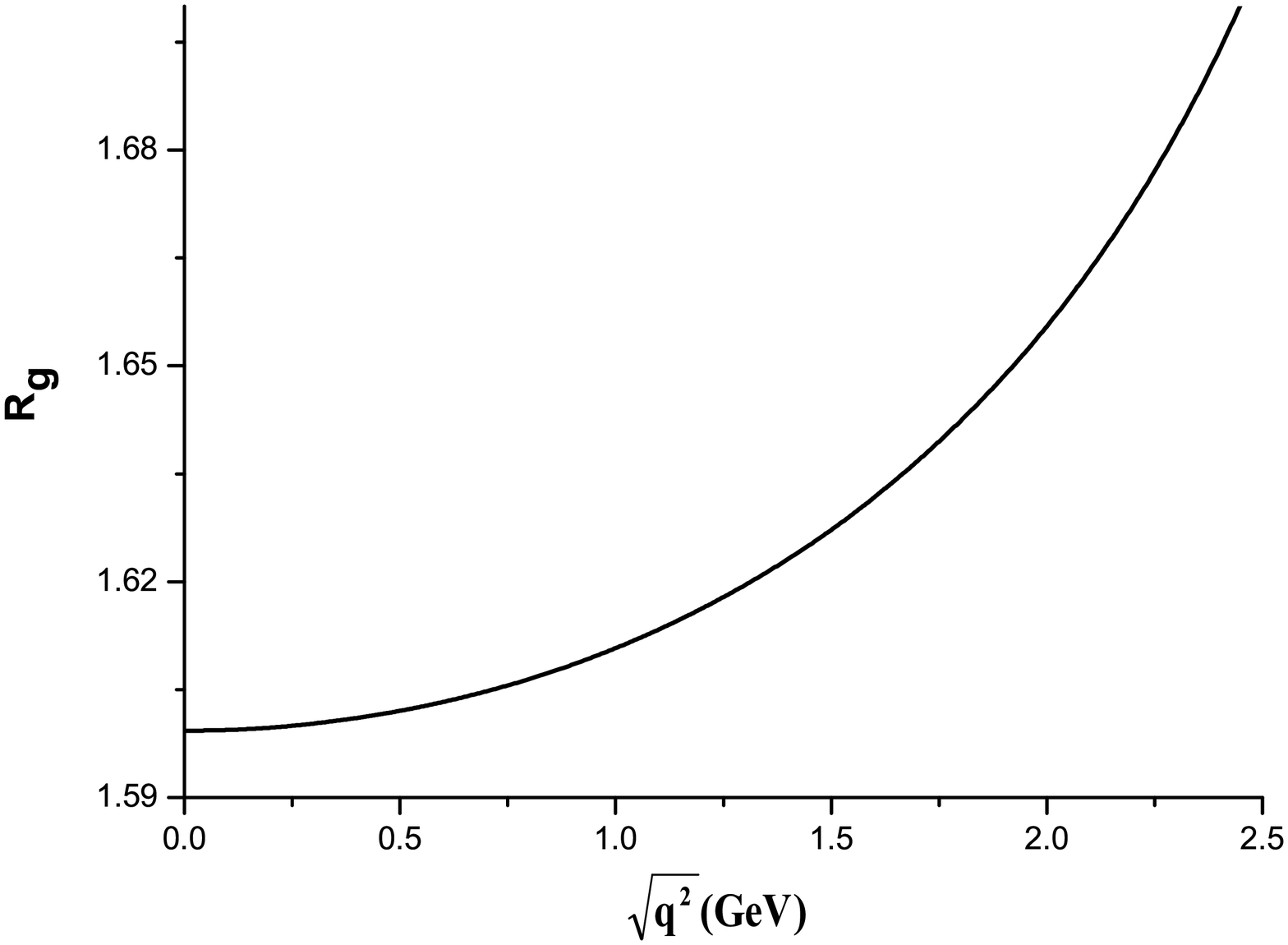}%
\includegraphics[width=0.5\textwidth]{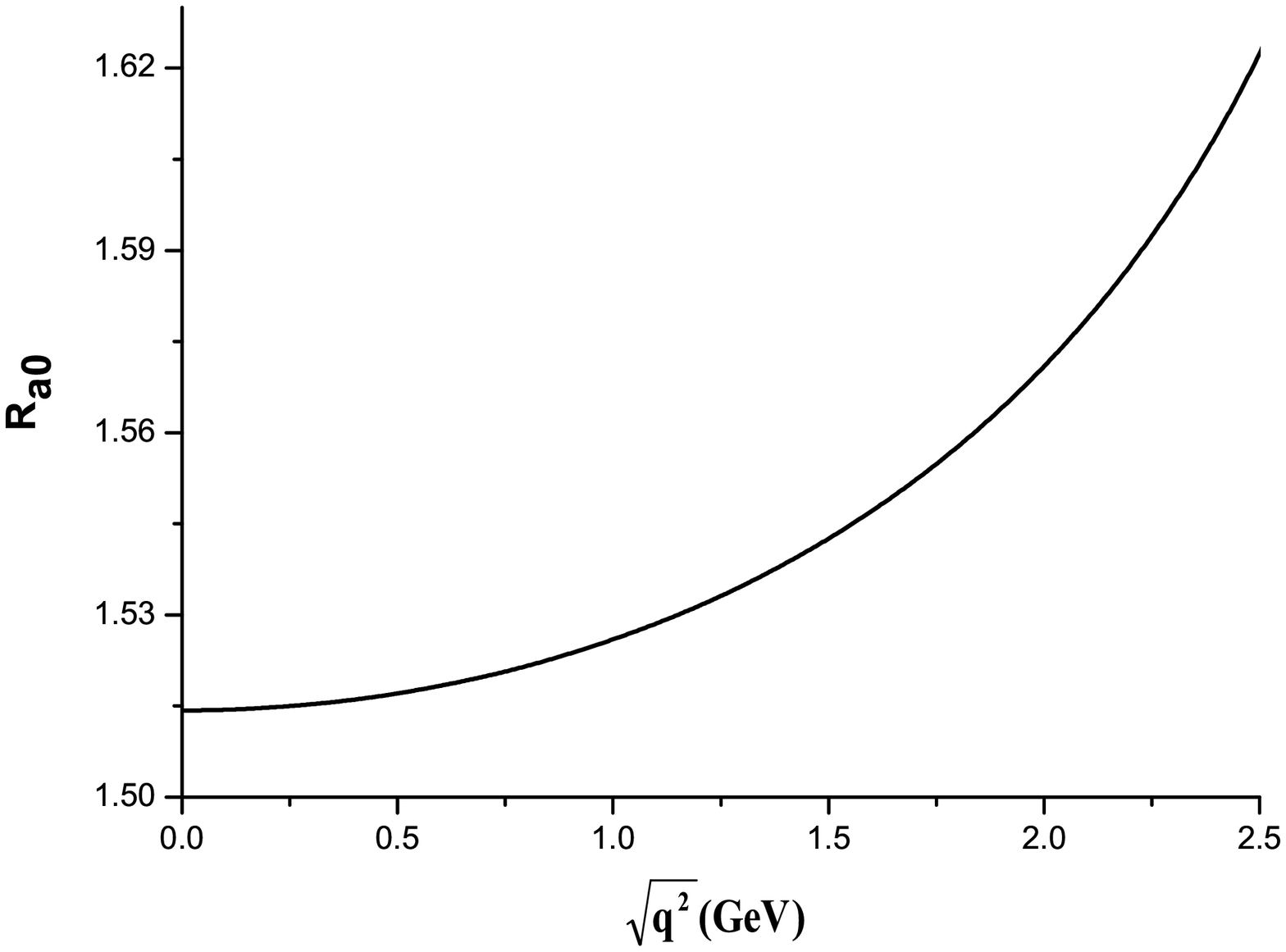}
\includegraphics[width=0.5\textwidth]{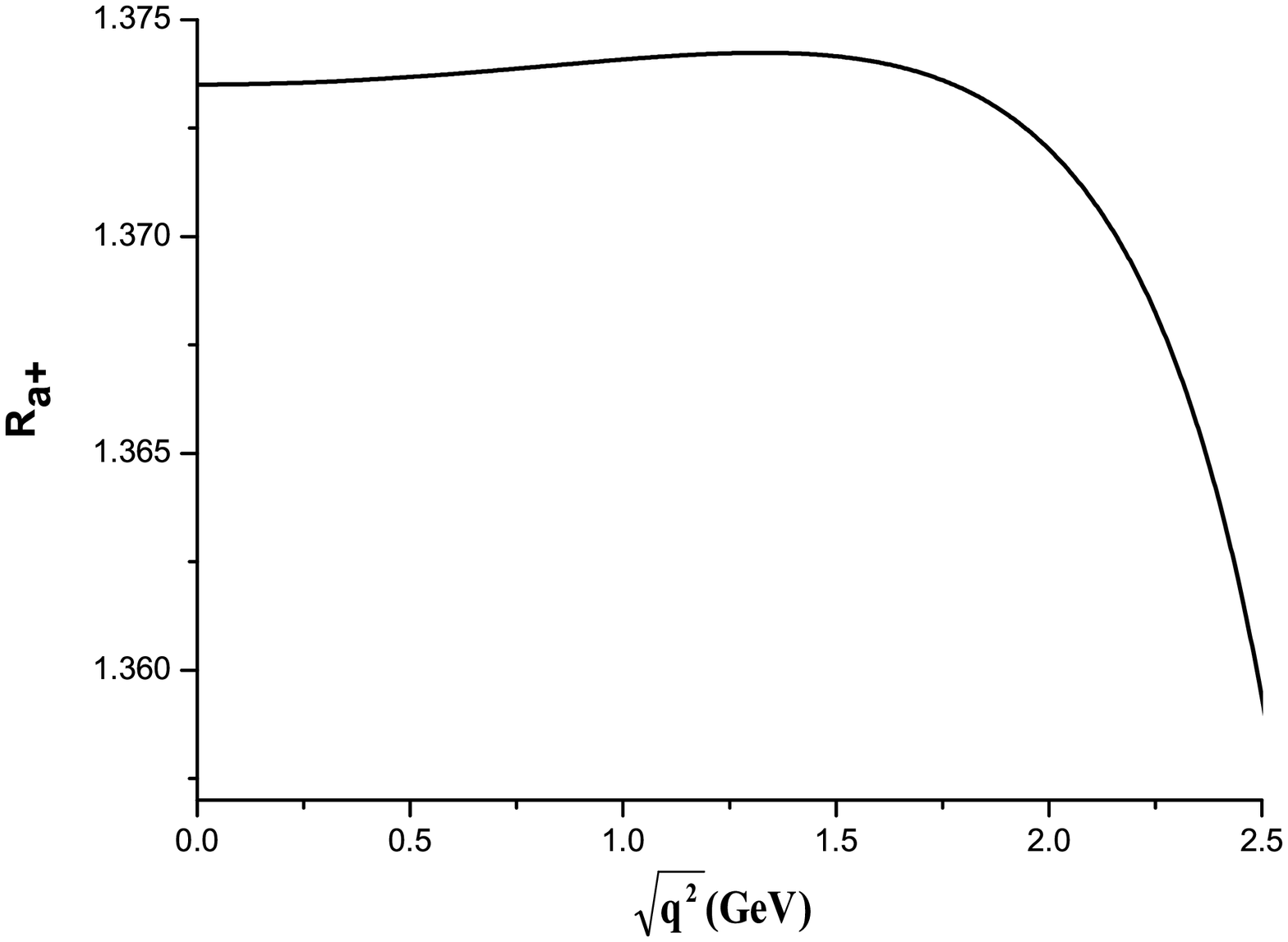}%
\includegraphics[width=0.5\textwidth]{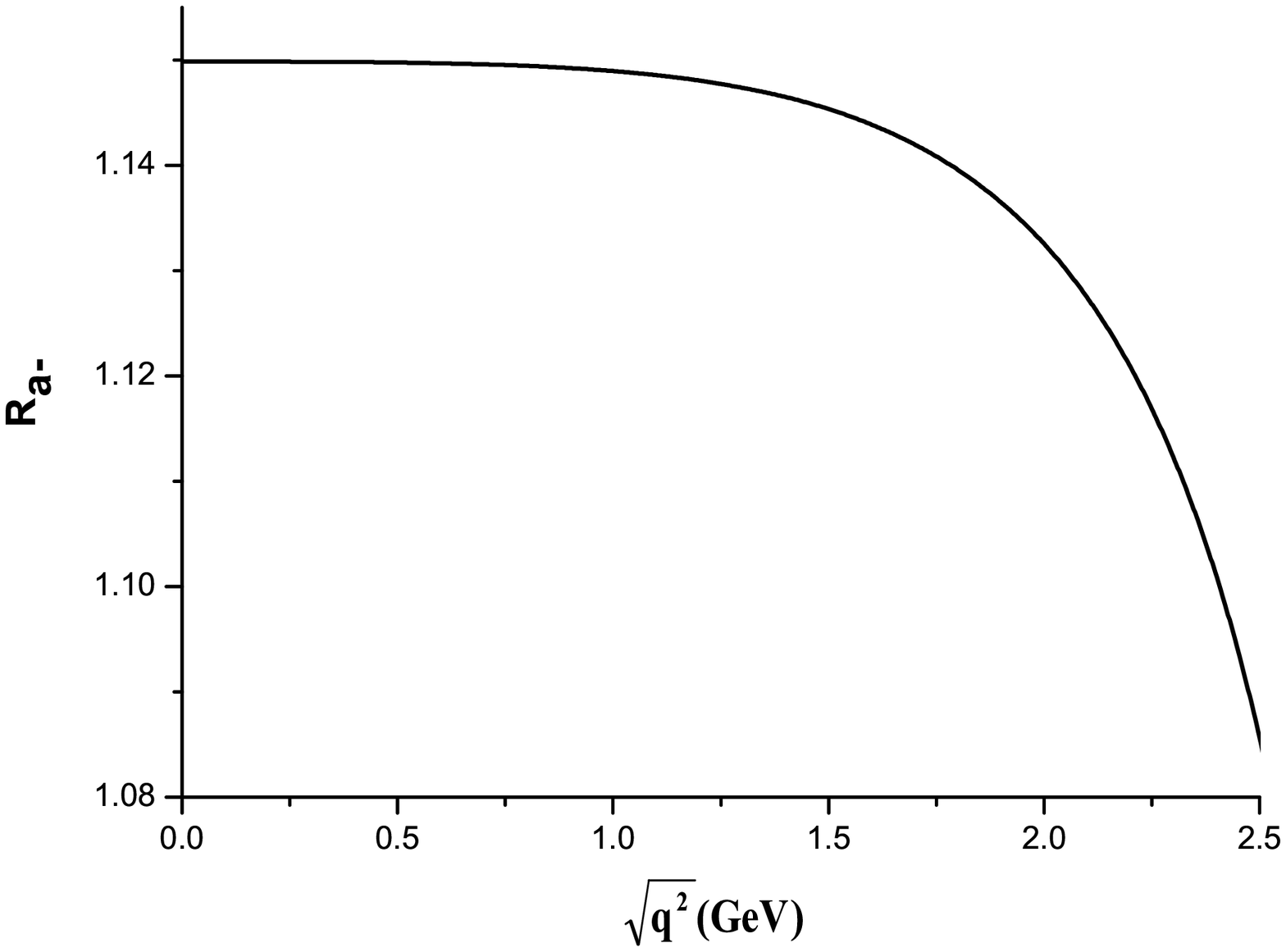}
 \caption{\small The ratios of NLO and LO form
factors vs the square root of momentum transfer $\sqrt{q^2}$. Here,
$\textmd{R}_i(q^2)= \frac{F^{NLO}_i(q^2)}{F_i^{LO}(q^2)}$ with $F_i$
standing for $f_{+}$, $f_{-}$, $g$, $a_{0}$, $a_{+}$, and $a_{-}$.
The renormalization-scale is fixed at $\mu=3$ GeV;  $m_{b}=4.76\;
\textmd{GeV}$ and $\; m_{c}=1.54\; \textmd{GeV}$ is adopted.}
\label{lpty} \vspace{-0mm}
\end{figure}

In performing the next-to-leading order calculation, as
schematically shown in Figure \ref{graph2}, we use the dimensional
regularization scheme to regularize the UV and IR divergences, and
the Coulomb divergence is regularized by the relative velocity $v$.
In dimensional regularization, it is well-known that the
$\gamma_{5}$ is difficult to deal with. In the literature, two
approaches are mostly employed, that is the Naive scheme
\cite{Chanowitz:1979zu} and the 't Hooft-Veltman scheme
\cite{'tHooft:1972fi}. In this calculation, we adopt the Naive
scheme, of which the $\gamma_{5}$ anticommutates with each
$\gamma^\mu$ matrix in d-dimension space-time,
$\{\gamma_{5},\gamma^{\mu}\}=0$. In evaluating the quarkonium
production and decays, it was argued by Ref. \cite{Petrelli:1997ge}
that both schemes may lead to the same result, which is different
from the case of pion decays to di-photon. The UV divergences exist
merely in self-energy and triangle diagrams, which can be
renormalized by the corresponding counter terms. The renormalization
constants include $Z_{2}$, $Z_{3}$, $Z_{m}$, and $Z_{g}$, referring
to quark field, gluon field, quark mass, and strong coupling
constant $\alpha_{s}$, respectively. In our calculation the $Z_{g}$
is defined in the modified-minimal-subtraction
($\mathrm{\overline{MS}}$) scheme, while for the other three the
on-shell ($\mathrm{OS}$) scheme is employed, which tells
\begin{eqnarray}
&&\hspace{-0.3cm}\delta Z_m^{OS}=-3C_F
\frac{\alpha_s}{4\pi}\left[\frac{1}{\epsilon_{UV}}-\gamma_{E}+
\ln\frac{4\pi\mu^{2}}{m^{2}}+\frac{4}{3}+{\mathcal{O}}(\epsilon)\right]\;
,\nonumber
\\ &&\hspace{-0.3cm}\delta Z_2^{OS}=-C_F
\frac{\alpha_s}{4\pi}\left[\frac{1}
{\epsilon_{UV}}+\frac{2}{\epsilon_{IR}}-3\gamma_{E}+3\ln
\frac{4\pi\mu^{2}}{m^{2}}+4+{\mathcal{O}}(\epsilon)\right]\; ,\nonumber\\
&&\hspace{-0.3cm}\delta Z_3^{OS}= \frac{\alpha_s}{4\pi}
\left[(\beta_0-2C_A)(\frac{1}{\epsilon_{UV}}-
\frac{1}{\epsilon_{IR}})+{\mathcal{O}}(\epsilon)\right]\; ,\nonumber\\
&&\hspace{-0.3cm}\delta Z_g^{\overline{MS}}=-\frac{\beta_0}{2}
\frac{\alpha_s}{4\pi}\left[\frac{1}{\epsilon_{UV}}-\gamma_{E}+
\ln4\pi+{\mathcal{O}}(\epsilon)\right]\; .\label{eq:13}
\end{eqnarray}
Here, $\beta_{0}=(11/3)C_{A}-(4/3)T_{f}n_{f}$ is the one-loop
coefficient of the QCD beta function; $n_{f}=3$ is the number of
active quarks in our calculation; $C_{A}=3$ and $T_{F}=1/2$
attribute to the SU(3) group; $\mu$ is the renormalization scale.

\begin{figure}[htb]
\centering\vspace*{-2cm}
\includegraphics[width=0.5\textwidth]{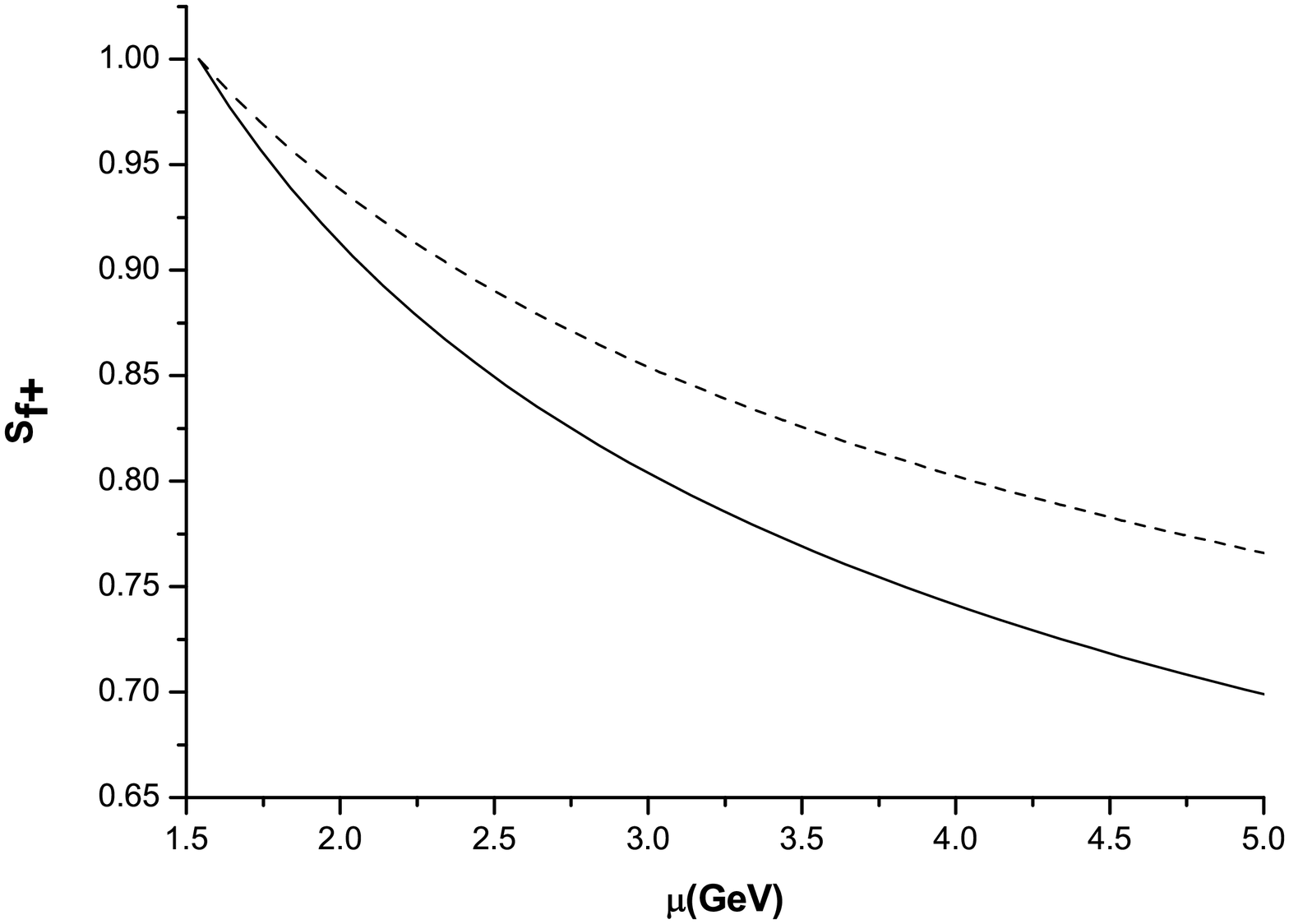}%
\includegraphics[width=0.5\textwidth]{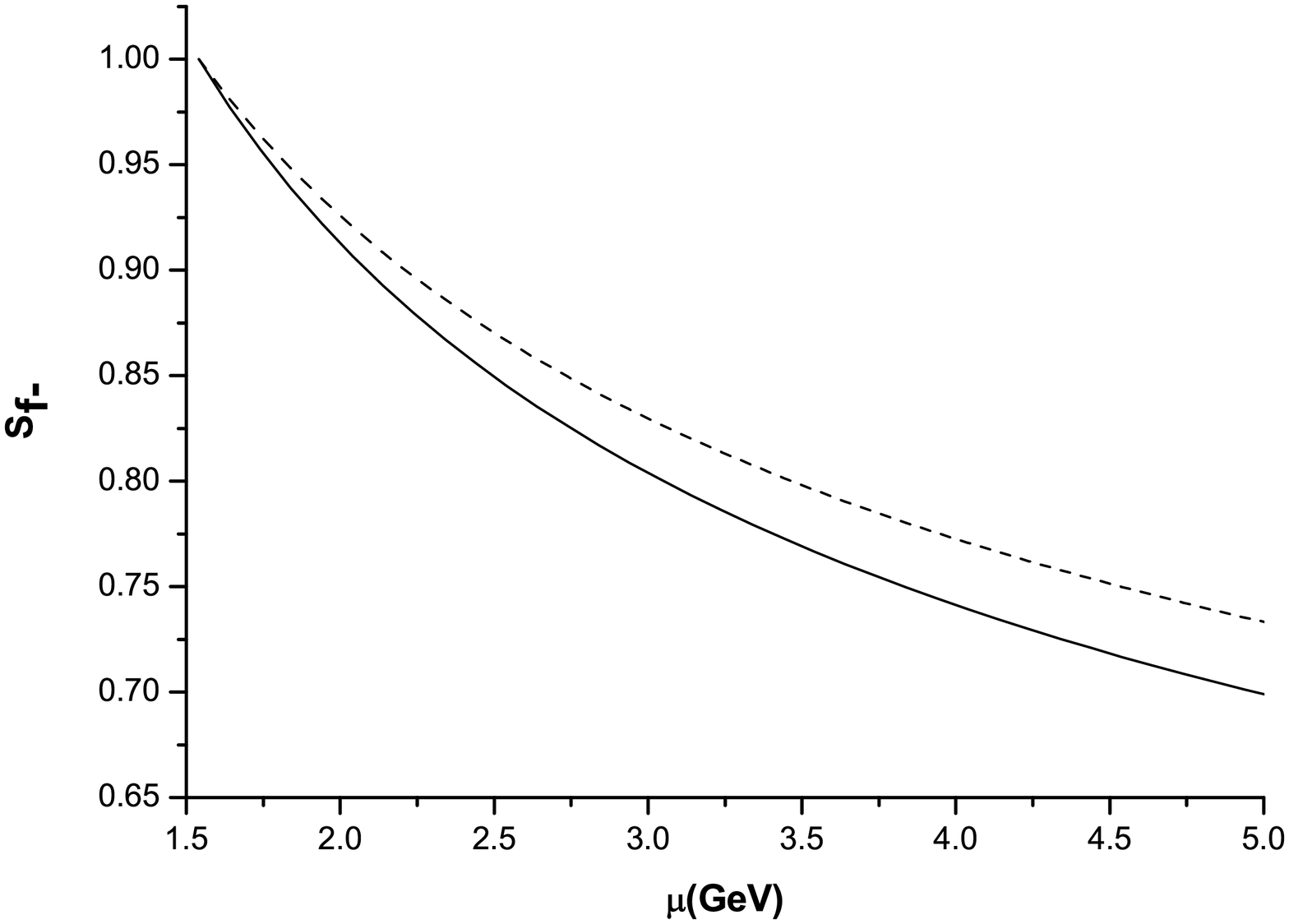}
\includegraphics[width=0.5\textwidth]{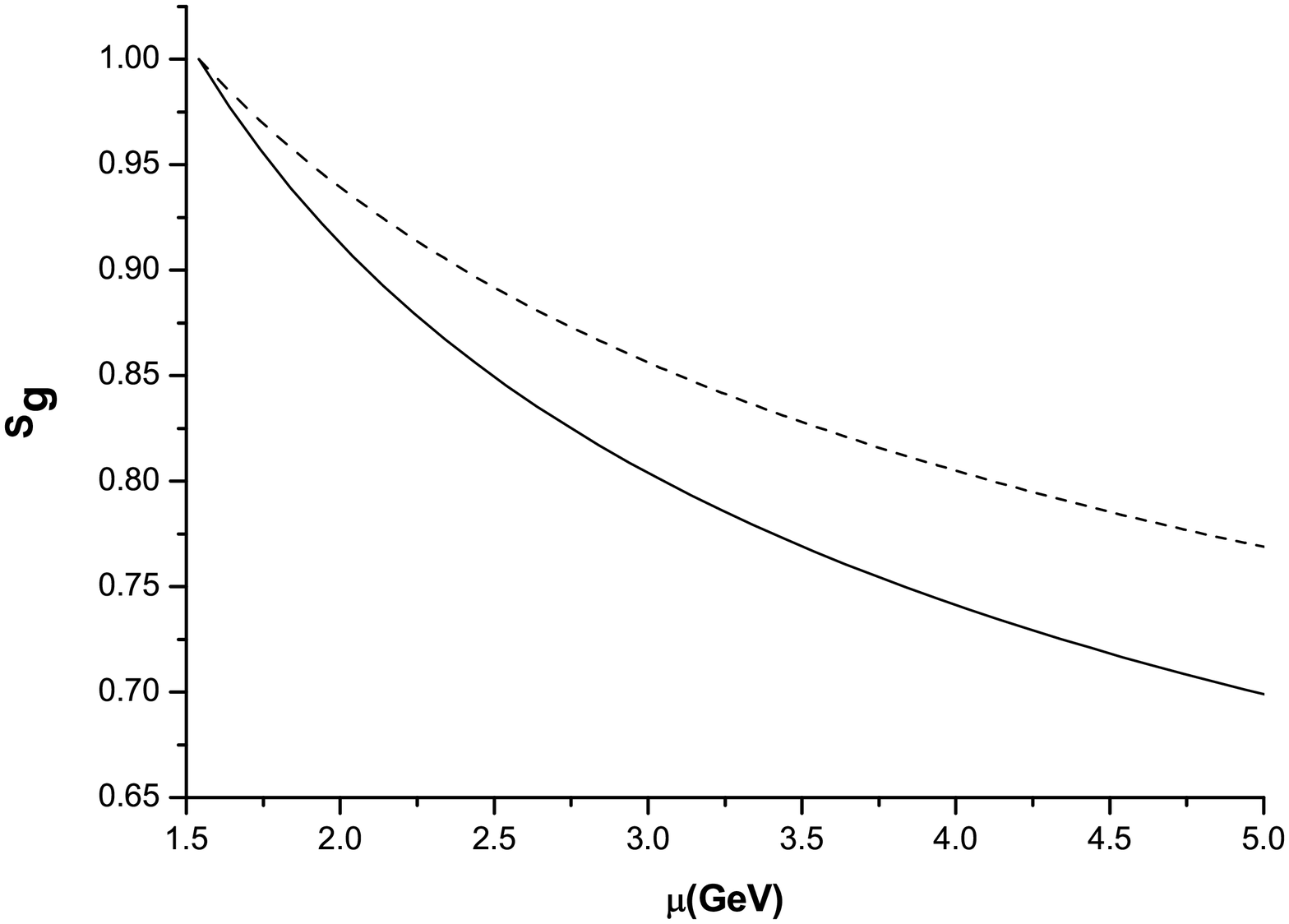}%
\includegraphics[width=0.5\textwidth]{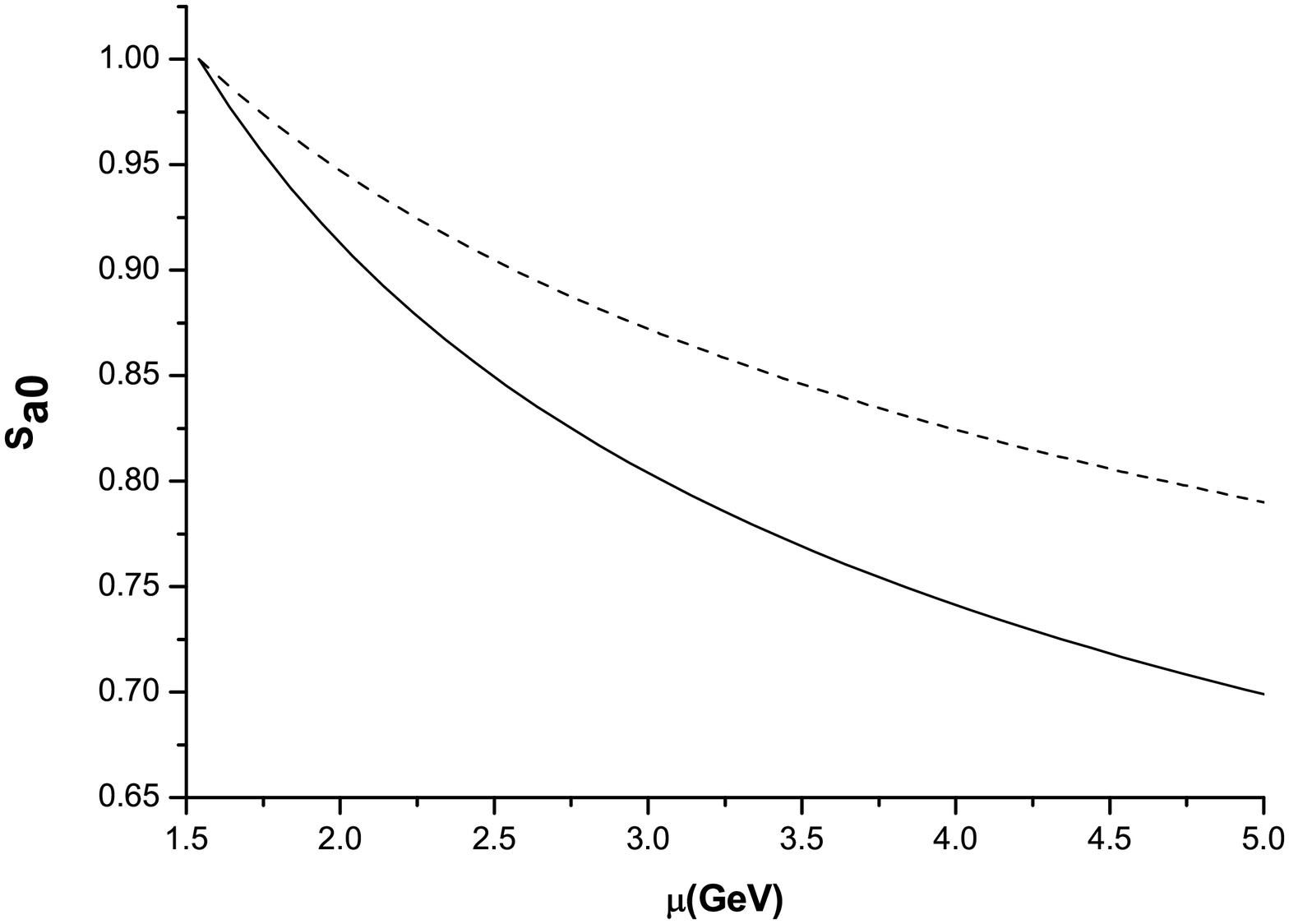}
\includegraphics[width=0.5\textwidth]{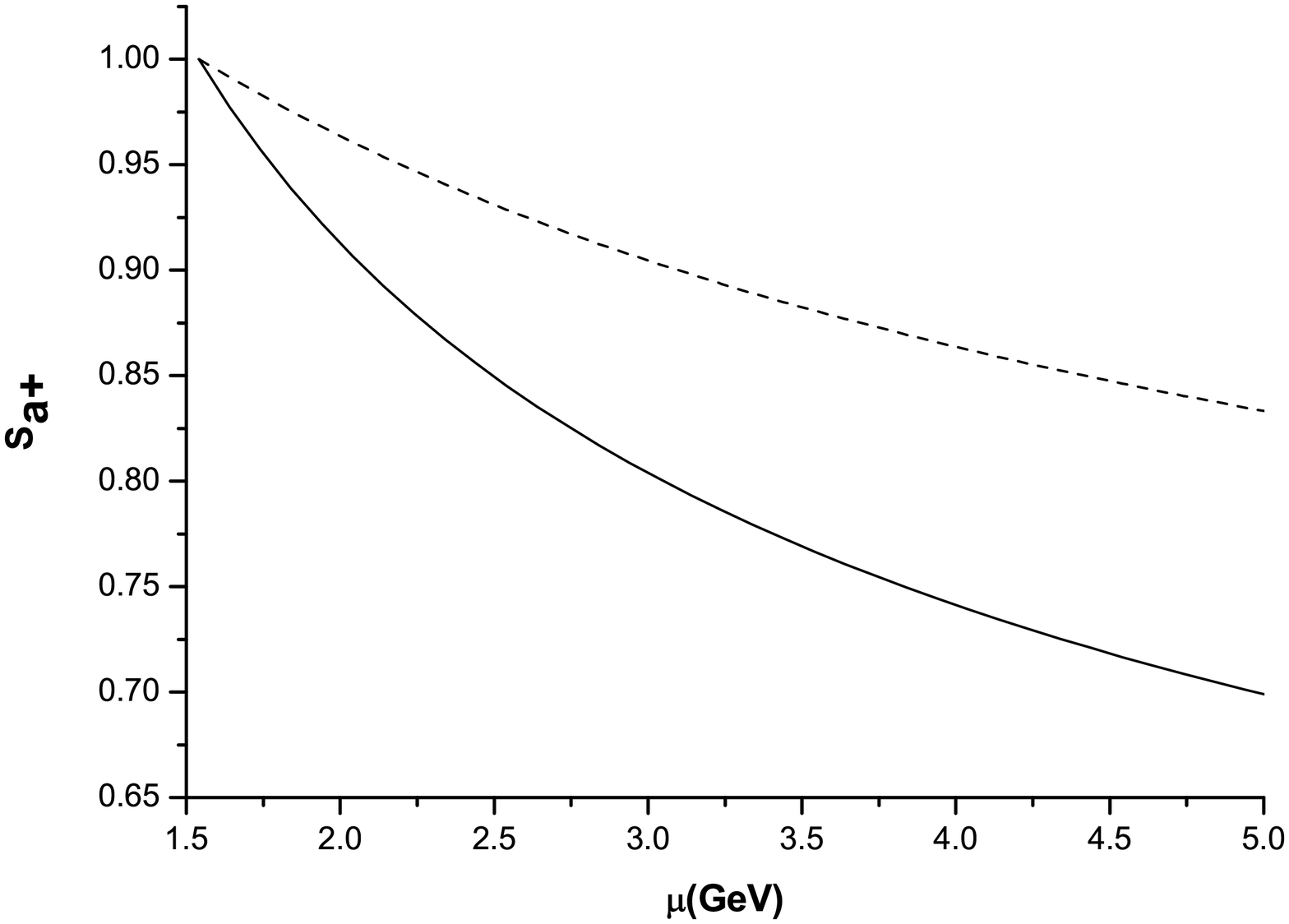}%
\includegraphics[width=0.5\textwidth]{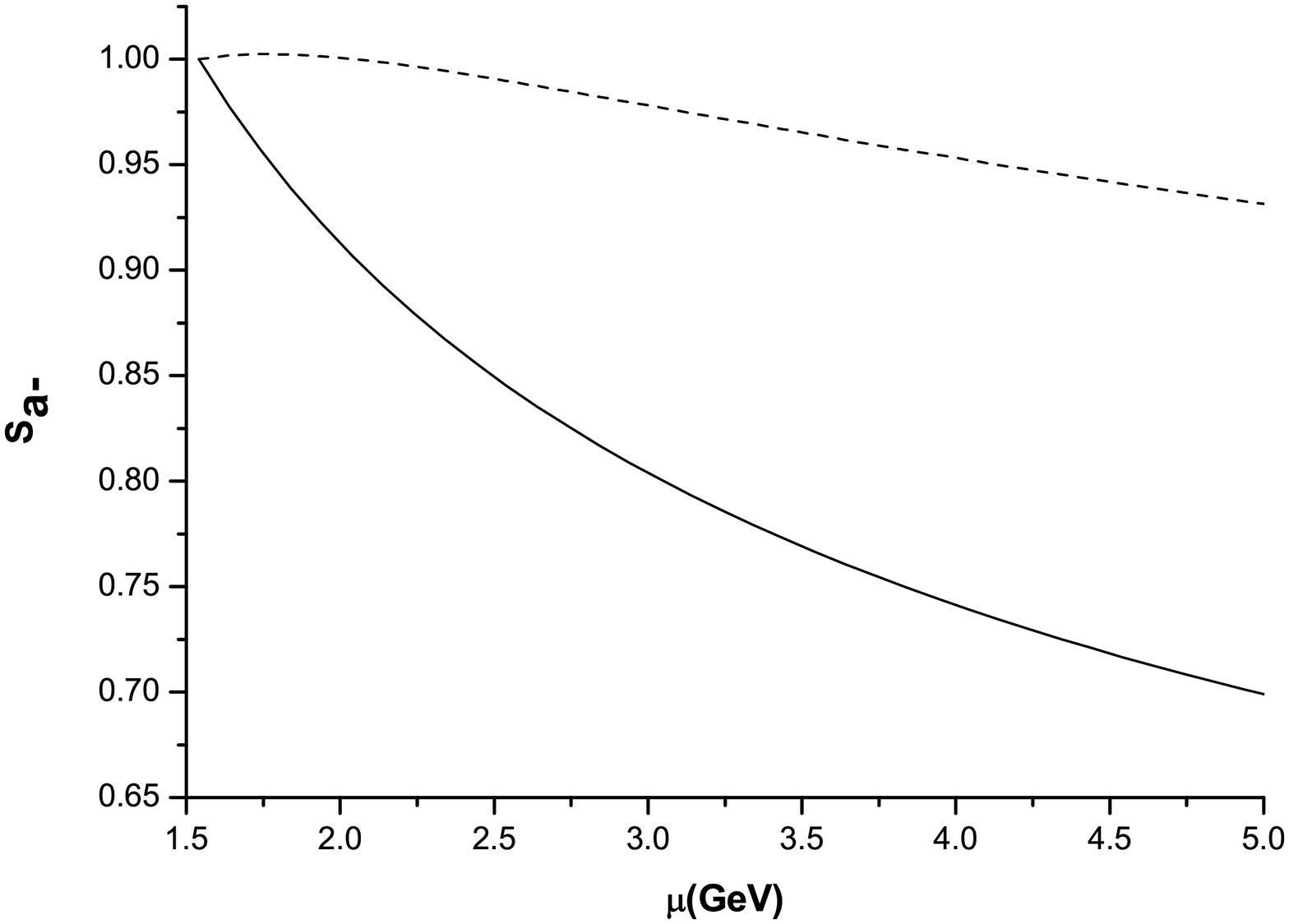}
 \caption{\small The renormalization-scale dependence of the LO and NLO form
 factors at the maximum recoil point, $q^2=0$. Here, $\textmd{S}_i(\mu)=
\frac{F_i(\mu)}{F_i(m_c)}$ with $F_i$ standing for $f_{+}$, $f_{-}$,
$g$, $a_{0}$, $a_{+}$, and $a_{-}$. The solid line represents the
situation of the LO renormalization-scale dependence and the dash
line represents for the NLO situation. In the computation,
$m_{b}=4.76\; \textmd{GeV}$ and $\; m_{c}=1.54\; \textmd{GeV}$ are
adopted.} \label{run1} \vspace{-0mm}
\end{figure}

Because in our calculation the $m_c/m_b$ contribution is kept, the
complete expression turns to be too lengthy to be presented here.
Whereas, the asymptotic form in small $m_c$ limit is given in the
appendix, and the numerical results are presented. From the
asymptotic form a noteworthy finding is that there exists an
interesting relation among those form factors, i.e,
\begin{eqnarray}
\frac{a_0^{NLO}}{a_0^{LO}}=\frac{a_+^{NLO}}{a_+^{LO}}
=\frac{g^{NLO}}{g^{LO}}\; ,\label{eq:14}
\end{eqnarray}
which is consistent with the prediction of Ref.\cite{Charles:1998dr}
from large energy effective theory(LEET).

In numerical calculation, the input heavy quark masses are
\begin{eqnarray}
m_{b}=4.76\; \textmd{GeV},\; m_{c}=1.54\; \textmd{GeV}\; .
\end{eqnarray}
The one loop result for strong coupling constant, the
\begin{eqnarray}
\alpha_s(\mu)=\frac{4\pi}{(11-\frac{2}{3}n_f)\mathrm{Log}(\frac{\mu^2}
              {\Lambda_{QCD}^2})}\; .\label{eq:24}
\end{eqnarray}
is used.

\begin{figure}[here]
\centering
\includegraphics[width=0.5\textwidth]{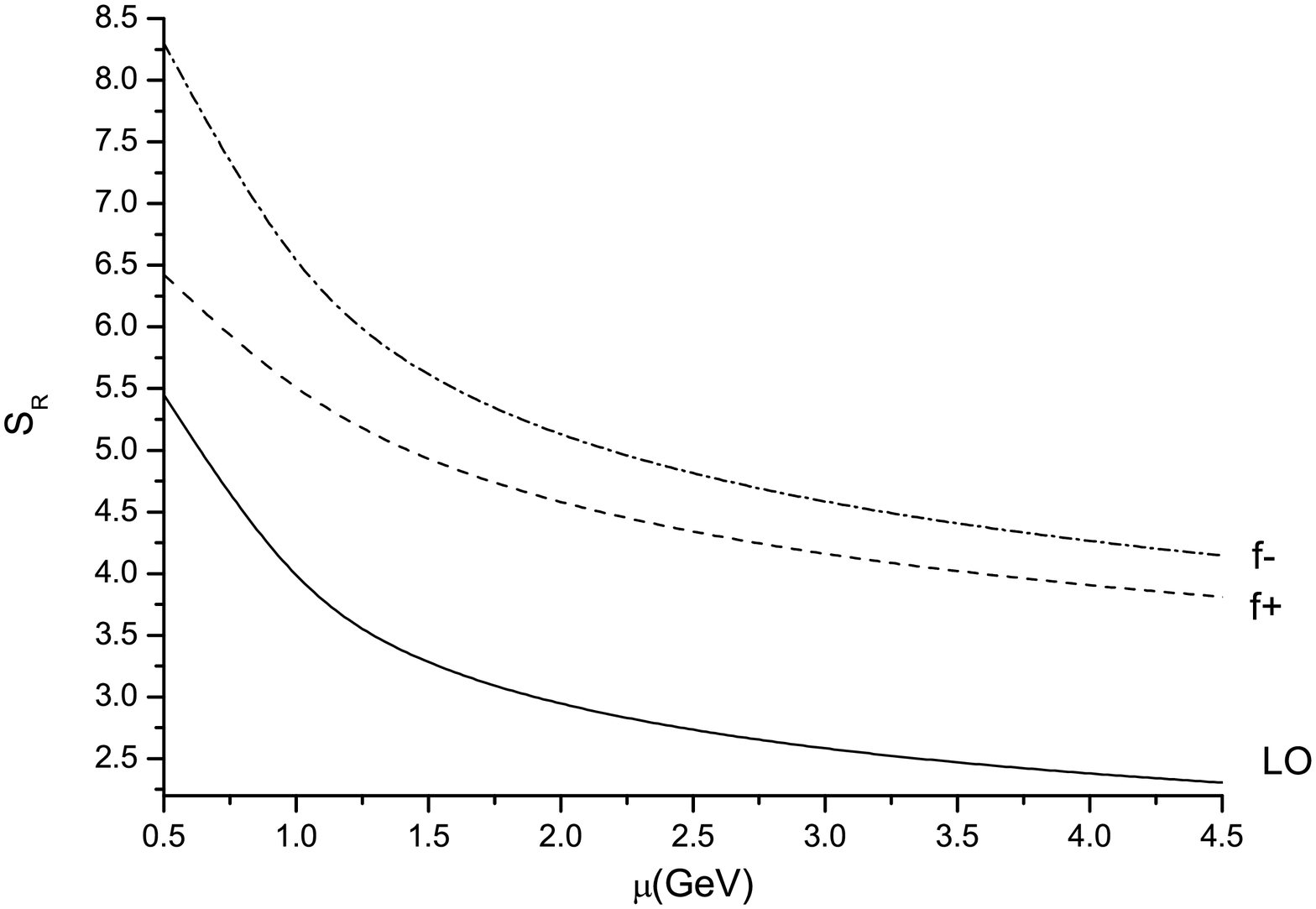}%
\includegraphics[width=0.5\textwidth]{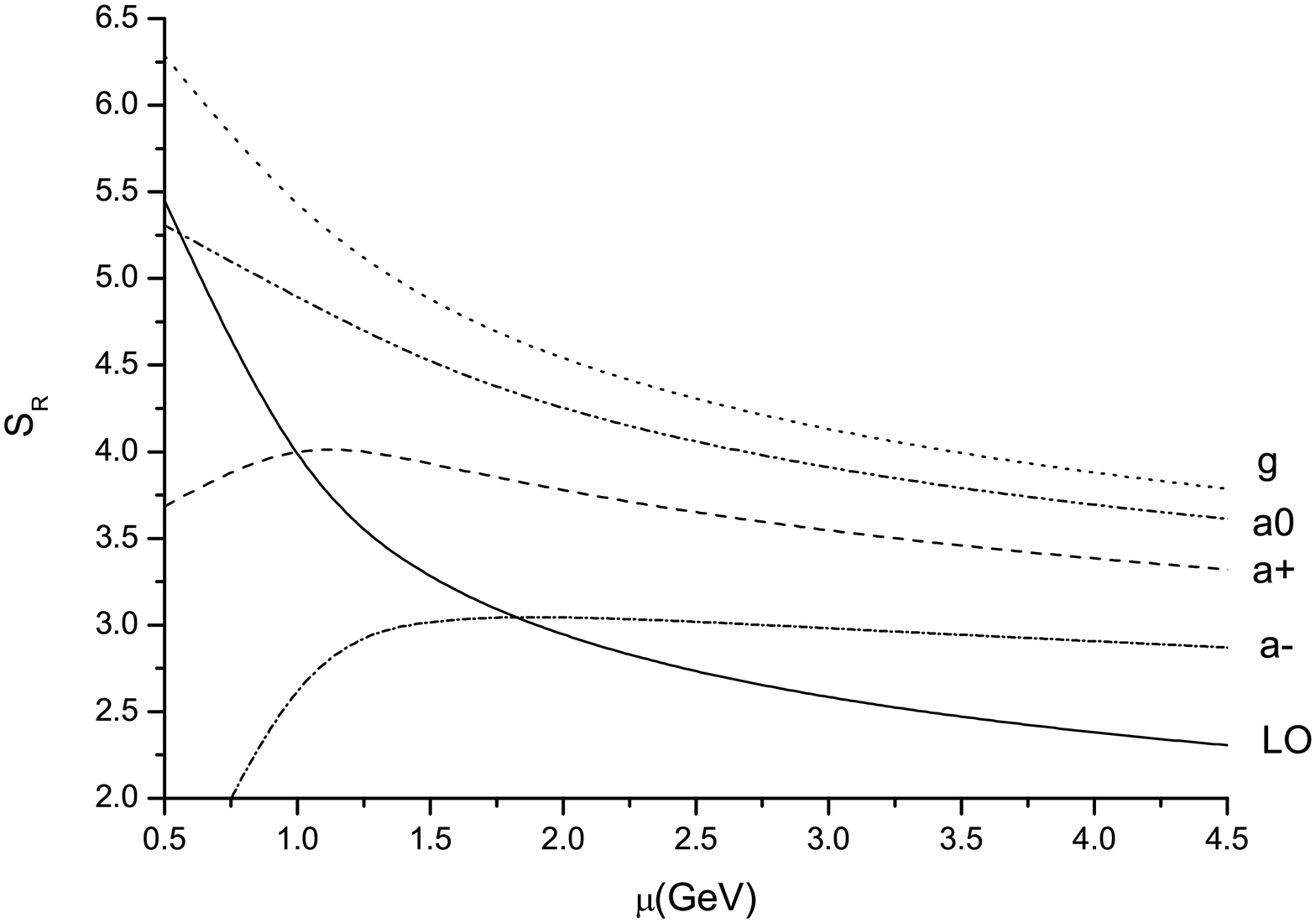}
 \caption{\small The renormalization-scale dependence of the LO and NLO form
 factors at the maximum recoil point, $q^2=0$. Here, $\textmd{S}^i_\textmd{R}(\mu)=
4\pi\alpha_s\frac{F_i(\mu)}{F_i^{LO}(\mu)}$ with $F_i$ standing for
$f_{+}$, $f_{-}$, $g$, $a_{0}$, $a_{+}$, and $a_{-}$. The symbol
``LO'' denotes the LO renormalization-scale running of
$\textmd{S}^i_\textmd{R}(\mu)$. In the computation, $m_{b}=4.76\;
\textmd{GeV}$ and $\; m_{c}=1.54\; \textmd{GeV}$ are adopted.}
\label{run11} \vspace{-0mm}
\end{figure}

\begin{figure}[here]
\centering
\includegraphics[width=0.5\textwidth]{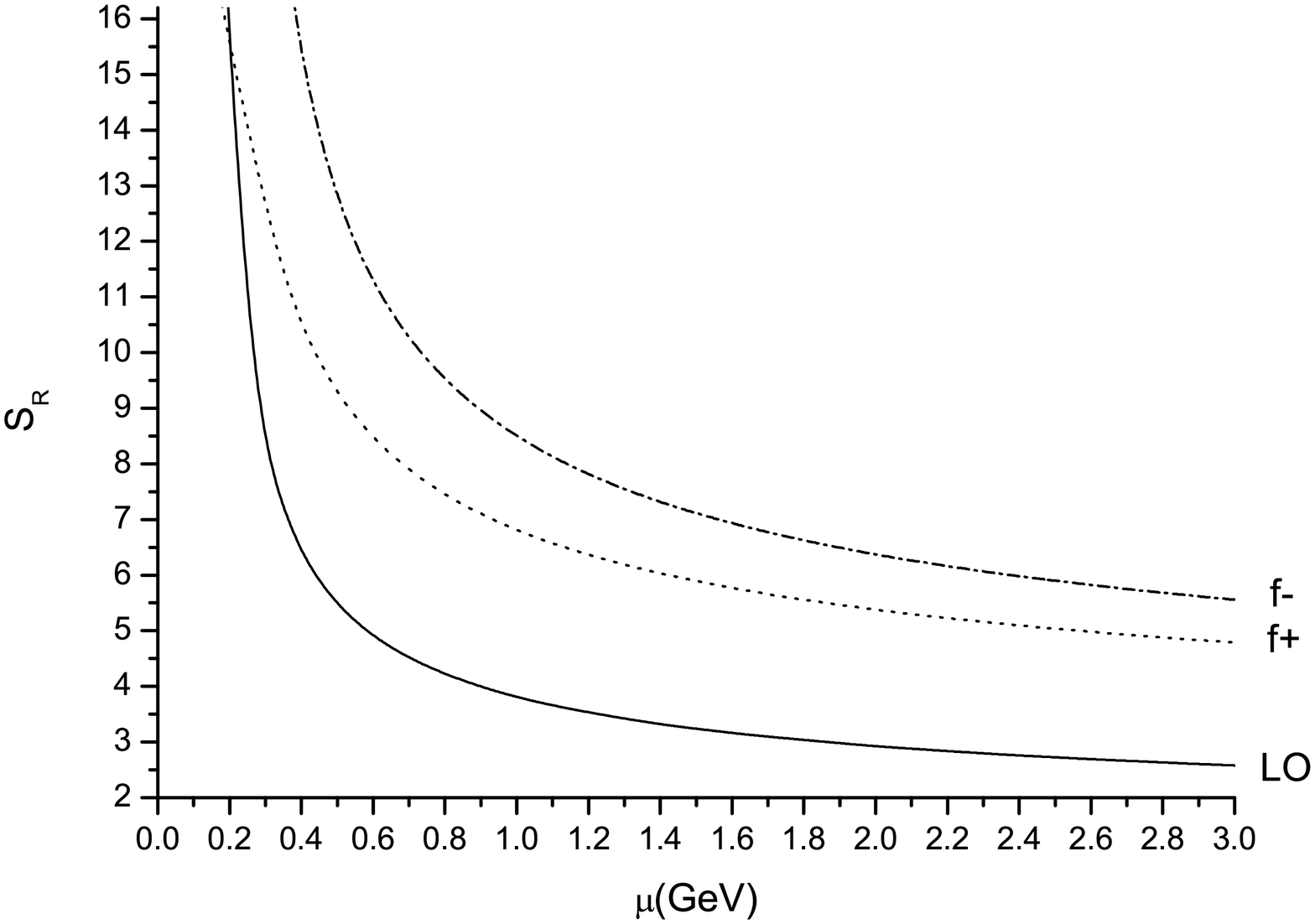}%
\includegraphics[width=0.5\textwidth]{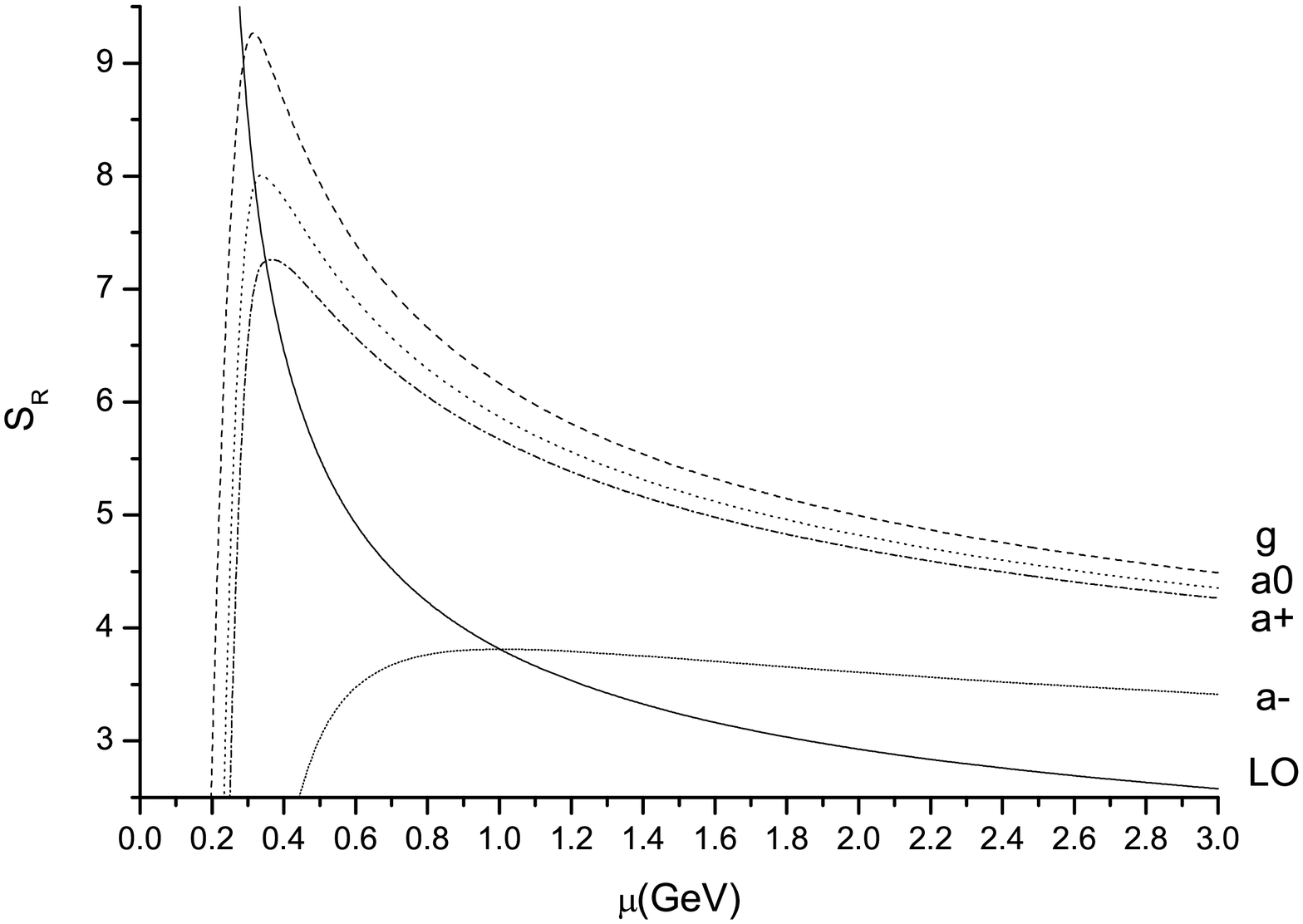}
 \caption{\small The renormalization-scale dependence of the LO and NLO form
 factors at the maximum recoil point, $q^2=0$.
 In the computation, $m_{b}=5\; \textmd{GeV}$ and
$m_{c}=0.3\; \textmd{GeV}$ are adopted and
$\textmd{S}^i_\textmd{R}(\mu)$ is performed to all orders of
$m_c/m_b$.} \label{run2} \vspace{-0mm}
\end{figure}

In Figure \ref{lpty}, the ratios of NLO and LO form factors versus
the square root of momentum transfer $\sqrt{q^2}$ are schematically
shown, while it should be noted that the pQCD approach is feasible
only in the maximum recoil region. The figure shows that the NLO
corrections to the $B_c$ to charmonia transition form factors are
remarkable and sensitive to the momentum transfer $q$. The
renormalization-scale dependence of the LO and NLO form factors are
presented in Figure \ref{run1} at the maximum recoil point $q^2=0$.
Generally speaking, the scale dependence in NLO is obviously
depressed relative to the LO case.

\begin{figure}[here]
\centering
\includegraphics[width=0.5\textwidth]{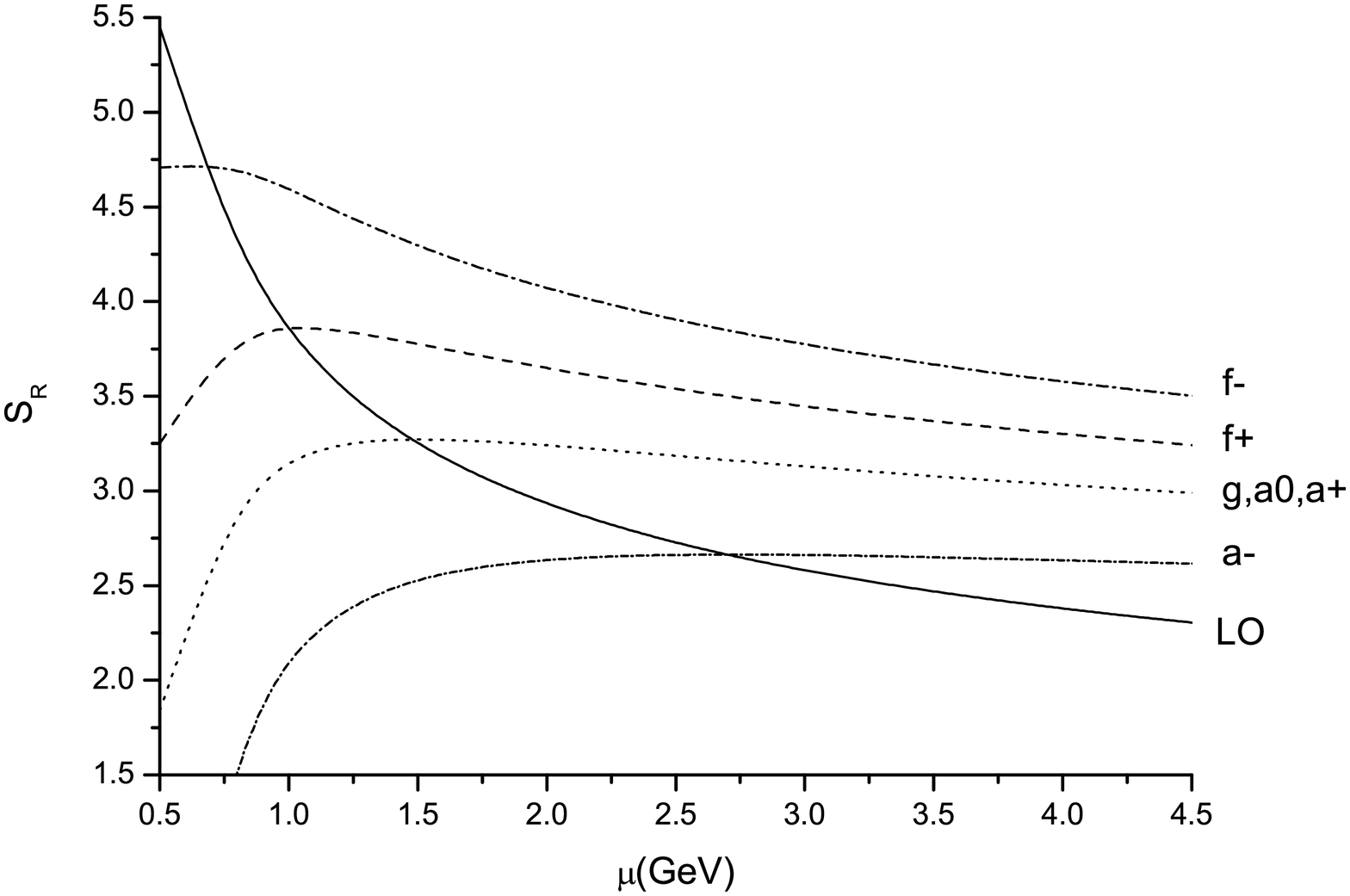}%
\includegraphics[width=0.5\textwidth]{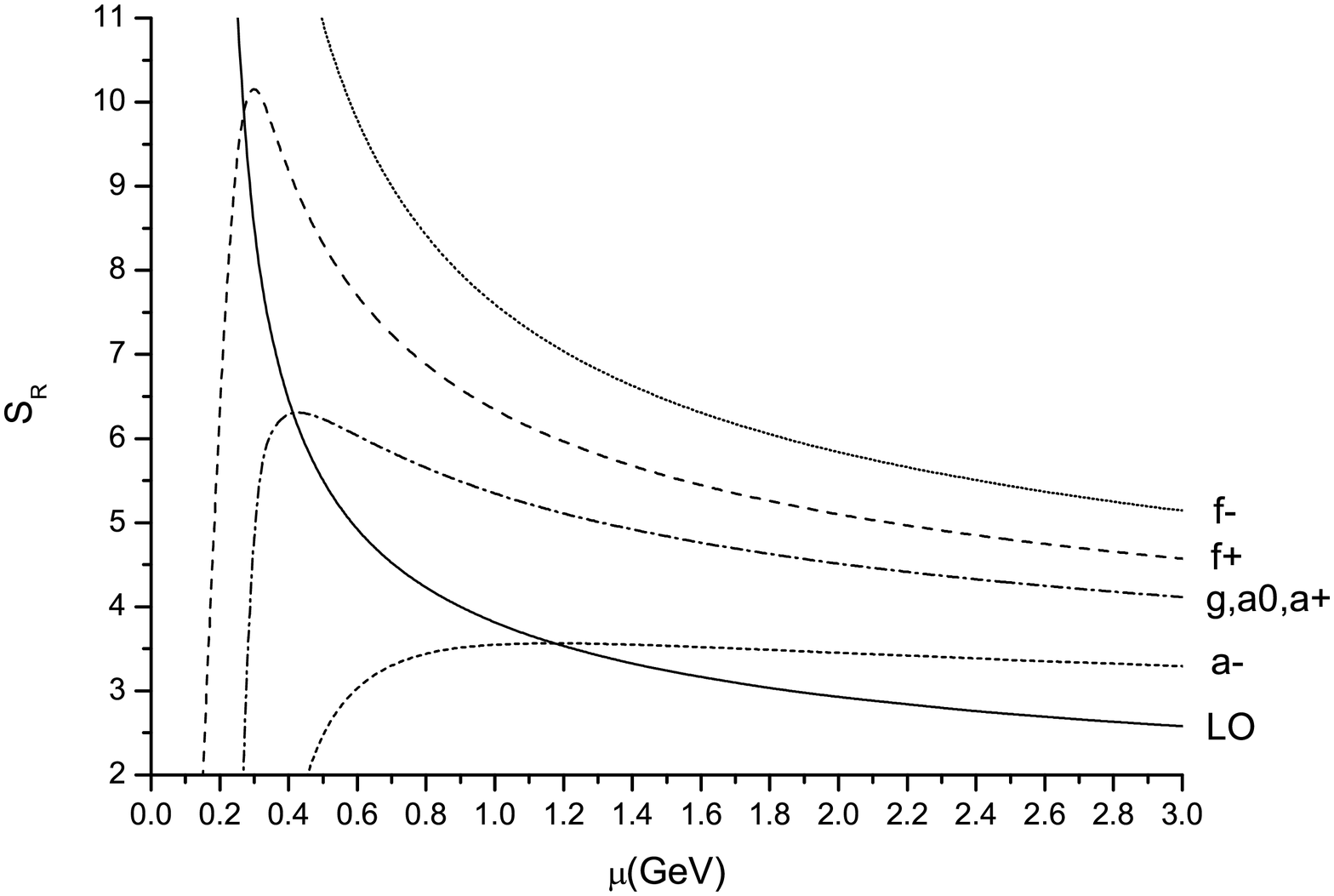}
 \caption{\small The renormalization-scale dependence of the LO and NLO form
 factors at the maximum recoil point, $q^2=0$.
 The left one is for $m_{b}=4.76\; \textmd{GeV}$ and
$m_{c}=1.54\; \textmd{GeV}$; the right one for $m_{b}=5\;
\textmd{GeV}$ and $m_{c}=0.3\; \textmd{GeV}$; and the
$\textmd{S}^i_\textmd{R}(\mu)$ is performed to the leading order of
$m_c/m_b$.} \label{run3} \vspace{-0mm}
\end{figure}

To show more explicitly the difference of LO and NLO results, we
employ the function of
$\textmd{S}^i_\textmd{R}(\mu)=4\pi\alpha_s(\mu)\frac{F_i(\mu)}{F_i^{LO}(\mu)}$
to exhibit the dependence of renormalization-scale at the maximum
recoil $q^2=0$, as shown in Figures \ref{run11} to \ref{run3}, where
$F_i(\mu)$ stands for $F_i^{LO}(\mu)$ or
$F_i^{LO}(\mu)+F_i^{NLO}(\mu)$ corresponding to LO
renormalization-scale dependence and NLO one respectively. To test
the validity of $m_c/m_b$ expansion, giving $m_c$ a nonphysical
small mass we present in Figures \ref{run2} and \ref{run3} the
leading order and full order results, respectively. From these
figures we see that those form factors are quite sensitive to the
magnitude of $m_c$, and the higher order effects in $m_c/m_b$
expansion are notable.

In our calculation the package FeynArts \cite{feynarts} was used to
generate the Feynman diagrams, the FeynCalc \cite{feyncalc} was used
to generated the amplitudes, and the LoopTools \cite{looptools} was
employed to calculate the Passarino-Veltman integrals.

\section{wave function overlap contribution}

The wave function overlap contribution for $B_c$ decays has been
broadly discussed in the literature
\cite{Du:1988ws,Colangelo:1992cx,Kiselev:1993ea,Kiselev:1999sc,
Nobes:2000pm,Ivanov:2000aj, Kiselev:2002vz,Ebert:2003cn,
Ivanov:2005fd,Hernandez:2006gt,Huang:2007kb,Sun:2008ew,DSV,Wang:2008xt}.
Since in the overlap contribution the QCD non-perturbative effects
tend to be important, from pQCD factorization point of view it is
beyond the scope of our interest in this work. However, here we
still make a schematic comparison of the wave-function overlap
contribution with pQCD contribution to the form factors for readers
convenience.

\begin{table}[here]
\caption{$B_c\to \eta_c$ and $J/\psi$ form factors at maximum recoil
$q^2=0$ from pQCD and wave function overlap \cite{Wang:2008xt}.
}\label{tab:formfactorcharmonium}
\begin{tabular}{c|cccccccccc}
 \hline\hline
    & $|F^{B_c \eta_c}|$ & $|A_0^{B_c J/\psi}|$ & $|A_1^{B_c J/\psi}|$
    & $|A_2^{B_c J/\psi}|$   & $|V^{B_c J/\psi}|$ \\
 \hline
    DW\cite{Du:1988ws}\footnote{We quote the results with $\omega=0.6$ GeV.}
    & $0.420$ & $0.408$ & $0.416$ & $0.431$  & $0.591$ \\
 \hline
    CNP\cite{Colangelo:1992cx}
     & $0.20$  & $0.26$ &$0.27$  & $0.28$  & $0.38$ \\
 \hline
    KT\cite{Kiselev:1993ea}
     & $0.23$  & $0.21$ &$0.21$  & $0.23$  & $0.33$ \\
 \hline
    KLO\cite{Kiselev:1999sc}\footnote{
    We quote the values where the Coulomb corrections are taken into account. }
     & $0.66$  & $0.60$ &$0.63$  & $0.69$  & $1.03$ \\\hline
    NW\cite{Nobes:2000pm}
     & $0.5359$ & $0.532$ &$0.524$  &$0.509$ & $0.736$\\
 \hline
    IKS\cite{Ivanov:2000aj}
     & $0.76$ & $0.69$  &$0.68$  & $0.66$ & $0.96$ \\
 \hline
    Kiselev\cite{Kiselev:2002vz}\footnote{The results
    out (in) the brackets are evaluated in sum rules (potential model). }
     & $0.66[0.7]$ & $0.60[0.66]$  &$0.63[0.66]$  & $0.69[0.66]$ & $1.03[0.94]$ \\
 \hline
    EFG\cite{Ebert:2003cn}
     & $0.47$ & $0.40$  &$0.50$  & $0.73$ & $0.49$ \\
 \hline
    IKS2\cite{Ivanov:2005fd}
     & $0.61$ & $0.57$  &$0.56$  & $0.54$ & $0.83$ \\
 \hline
    HNV\cite{Hernandez:2006gt}
     & $0.49$ & $0.45$  &$0.49$  & $0.56$ & $0.61$ \\
 \hline
    HZ\cite{Huang:2007kb}
     & $0.87$ &  $0.27$ &$0.75$  & $1.69$ & $1.69$ \\
    \hline
    SDY\cite{Sun:2008ew}
     & $0.87$ &  $0.27$ &$0.75$  & $1.69$ & $1.69$ \\
     \hline
    DSV\cite{DSV}
     & $0.58$ &  $0.58$ &$0.63$  & $0.74$ & $0.91$ \\
     \hline
      WSL\cite{Wang:2008xt}
     & $0.61$ &  $0.53$ &$0.50$  & $0.44$ & $0.74$ \\
     \hline
   pQCD(LO)\footnote{$\psi_{B_c}(0)=0.36\ \textmd{GeV}^{3/2}$,\
   $\psi_{J/\psi}(0)=0.26\ \textmd{GeV}^{3/2}$,\ $\alpha_s=0.2$ }
     & $0.97$ &  $0.86$ &$0.90$  & $0.97$ & $0.32$ \\
      \hline
  pQCD(NLO)$^d$
     & $1.57$ &  $1.34$ &$1.35$  & $1.39$ & $0.52$ \\
    \hline\hline
\end{tabular}
\end{table}

From the Table I, we notice that the results from wave function
overlap are comparable to what from the pQCD calculation at the
maximum recoil point, though in fact they attribute to different
mechanisms in the study of $B_c$ decays.

\section{Summary and Conclusions}

In this work we have calculated the transition form factors of the
$B_c$ meson to S-wave charmonia in the degree of NLO accuracy of
pQCD. In our calculation, the $B_c$ meson and charmonia are treated
as non-relativistic bound states of two heavy quarks. Hence, the
long-distance effects for the form factors come only from the soft
gluon exchange between heavy quarks, which can be explicitly
factorized out and expressed as the wave functions at the origin.
The factorization scale is set to be at $vm_{c}$ in practical
calculation, although the scale dependence does not appear at the
one-loop level. Since in the small recoil region the end-point
divergence spoils the QCD factorization, our result is valid only in
the large recoil case.

Calculation shows that the NLO QCD corrections to the $B_c$ to
charmonia form factors are remarkable, especially for the $f_-$
where the correction is as large as 80\% or so. We find that the
renormalization-scale dependence of the form factors are depressed,
as it should be, when the next-to-leading order correction is taken
into account, which means the uncertainties in the theoretical
estimation are reduced. We find that $B_c$ to charmonia transition
form factors are quite sensitive to the magnitude of $m_c$, and the
higher order effects in $m_c/m_b$ expansion are indispensable.

Last, it should be mentioned that the relativistic corrections are
also important for the $B_c$ to charmonia form factors, which are
highly related to the relative velocity of heavy quarks within the
bound states, i.e. $v^2_{\bar{b}c}\sim 0.38$ and $v^2_{\bar{c}c}\sim
0.25$ for $B_c$ and charmonium, respectively.

\vspace{.0cm} {\bf Acknowledgments}

We thank De-Shan Yang and Rui-Lin Zhu for useful discussion and
independent check of our result. This work was supported in part by
the National Natural Science Foundation of China(NSFC) under the
grants 10935012, 10928510, 10821063 and 10775179, by the CAS Key
Projects KJCX2-yw-N29 and H92A0200S2.


\vspace{.7cm} \noindent {\bf \large Appendix} \vspace{.3cm}

Following, various $B_c$ to charmonium transition form factors are
given in leading power of $m_c/m_b$ and NLO pQCD. For the sake of
compactness, we define $s=\frac{m_b^2}{m_b^2-q^2}$ and
$\gamma=\frac{m_b^2-q^2}{4m_bm_c}$. It is worth emphasizing that our
expressions for $\frac{f^{NLO}_+}{f^{LO}_+}$ and
$\frac{f^{NLO}_-}{f^{LO}_-}$ agree with what given in reference
\cite{Bc2}.
\begin{eqnarray}
\frac{f^{NLO}_+(q^2)}{f^{LO}_+(q^2)}&=&1+\frac{\alpha_s}{4\pi}\{\frac{1}{3}
(11 C_A-2 \text{n}_f) \log
 (\frac{\mu^2}{2 \gamma\text{m}_c^2})-\frac{10\text{n}_f}{9}+\frac{(\pi
^2-6 \log (2)) (s-1)+3 s \log (\gamma )}{6 s+3}\nonumber\\
&&+\frac{C_A}{72 s^2-18}(18 s^2 (2 s-1) \log ^2(s)+18 (8 \log (2)
s^3-2 \log (2) s^2-5 \log (2) s+s\nonumber\\
&&+2 \log (2)) \log (s)+(2 s-1) (268 s+\pi ^2 (6 s^2-3
   s-6)+170)-9 (2 s\nonumber\\
&&-1) \log (\gamma ) (\log (\gamma ) s-(2+2 \log (2)) s+4 \log
(2))+18 (2 s-1) (4 s^2+s\nonumber\\
&&-2) \text{Li}_2(1-2 s)-18 (4 s^3-5
   s+2) \text{Li}_2(1-s)+18 (s (4 s (s+1)-11)\nonumber\\
&&+4) \log ^2(2)-36 (5 (s-1) s+1) \log
   (2))\nonumber\\
&&+\frac{C_F}{6 (1-2 s)^2 (2 s+1)}(-6 (2 (s-1) s-1) \log ^2(s) (1-2
s)^2+3 \log (\gamma ) (23 s\nonumber\\
&&+(5 s-2) \log (\gamma )-4 (s+1) \log (2)+12) (1-2 s)^2-12 (4
s^2+s\nonumber\\
&&-2) \text{Li}_2(1-2 s)
   (1-2 s)^2+12 (s (2 s+3)-1) \text{Li}_2(1-s) (1-2 s)^2\nonumber\\
&&-(\pi -2 \pi  s)^2 (s (4 s-19)+4)+3 (-32 \log ^2(2) s^4-4 (69+2
\log (2) (-37\nonumber\\
&&+5 \log (2))) s^3+8 (18+\log
   (2) (-31+9\log (2))) s^2+(61+28 \log (2)\nonumber\\
&&-26 \log ^2(2)) s+12\log (2)+2 \log ^2(2)-32)+(6 s (8 s (s (-4
\log (2) s\nonumber\\
&&+3 \log (2)+3)+2 \log (2)-3)-18
   \log (2)+7)+24 \log (2)) \log (s))\}
\end{eqnarray}
\begin{eqnarray}
\frac{f^{NLO}_+(0)}{f^{LO}_+(0)}&=&1+\frac{\alpha_s}{4\pi}\{\frac{1}{3}
(11 C_A-2 \text{n}_f) \log
 \left(\frac{2\mu^2}{m_bm_c}\right)-\frac{10\text{n}_f}{9}+\frac{1}{3} \log
 \left(\frac{m_b}{m_c}\right)-\frac{2 \log (2)}{3}\nonumber\\
&&+C_F\left(\frac{1}{2} \log
^2\left(\frac{m_b}{m_c}\right)-\frac{10}{3} \log (2) \log
\left(\frac{m_b}{m_c}\right)+\frac{35}{6} \log
\left(\frac{m_b}{m_c}\right)+\frac{2 \log
   ^2(2)}{3}\right.\nonumber\\
&&\left.+3 \log (2)+\frac{7 \pi ^2}{9}-\frac{103}{6}\right)\nonumber\\
&&+C_A\left(-\frac{1}{6} \log
^2\left(\frac{m_b}{m_c}\right)+\frac{1}{3} \log (2) \log
\left(\frac{m_b}{m_c}\right)+\frac{1}{3} \log
\left(\frac{m_b}{m_c}\right)+\frac{\log
   ^2(2)}{3}\right.\nonumber\\
&&\left.-\frac{4 \log (2)}{3}-\frac{5 \pi
^2}{36}+\frac{73}{9}\right)
\end{eqnarray}
\begin{eqnarray}
\frac{f^{NLO}_-(q^2)}{f^{LO}_-(q^2)}&=&1+\frac{\alpha_s}{4\pi}\{\frac{1}{3}
(11 C_A-2 \text{n}_f) \log
 (\frac{\mu^2}{2 \gamma\text{m}_c^2})-\frac{10\text{n}_f}{9}+\frac{1}{6}
 (3 \log (\gamma )-6 \log (2)+\pi ^2)\nonumber\\
&&+\frac{C_A}{36 (s-1) (2 s-1)}(18 (s-1) s (2 s-1) \log ^2(s)+18 ((2
s
(4 s-5)+1) \log (2) s\nonumber\\
&&+s+\log (2)+1) \log (s)+(s-1) (2 s-1) (\pi ^2 (6 s-3)+268)\nonumber\\
&&+9 (s-1) (2 s-1) (-\log (\gamma
   )+2 \log (2)+2) \log (\gamma )+18 (8 s^3-10 s^2+s\nonumber\\
&&+1) \text{Li}_2(1-2 s)-18 (s-1) (2 s-1) (2 s+1)
\text{Li}_2(1-s)+18 (4 s^3-7 s\nonumber\\
&&+3) \log
   ^2(2)-36 (s-1) (5 s-1) \log (2))\nonumber\\
&&-\frac{C_F}{12 (1-2 s)^2 (s-1)^2}(12 (1-2 s)^2 \log ^2(s)
(s-1)^3+(\pi ^2 (s-1) (4 s-19) (1-2 s)^2\nonumber\\
&&+3 (32 \log ^2(2) s^4+4 (69+2 (-37+\log (2)) \log (2)) s^3+(-508-8
\log (2) (-74\nonumber\\
&&+13\log
   (2))) s^2+(307-364 \log (2)+82 \log ^2(2)) s+2 (34-9
   \log (2)) \log (2)\nonumber\\
&&-61)) (s-1)+6 (s (s (-24 s^2+84 s+2 (2 s-1) (2 s (4
   s-9)+11) \log (2)-127)\nonumber\\
&&-4 \log (2)+73)+2 \log (2)-13) \log (s)+3 (2 s^2-3 s+1)^2 (-5 \log
(\gamma )+4 \log (2)\nonumber\\
&&-23) \log (\gamma )+12 (4
   s+1) (2 s^2-3 s+1)^2 \text{Li}_2(1-2 s)\nonumber\\
&&-12 (2 s+3) (2 s^2-3 s+1)^2
   \text{Li}_2(1-s))
\end{eqnarray}
\begin{eqnarray}
\frac{f^{NLO}_-(0)}{f^{LO}_-(0)}&=&1+\frac{\alpha_s}{4\pi}\{\frac{1}{3}
(11 C_A-2 \text{n}_f) \log
 \left(\frac{2\mu^2}{m_bm_c}\right)-\frac{10\text{n}_f}{9}+\frac{1}{2} \log
 \left(\frac{m_b}{m_c}\right)-2 \log (2)+\frac{\pi ^2}{6}\nonumber\\
&&+C_F\left(\frac{5}{4} \log ^2\left(\frac{m_b}{m_c}\right)-6 \log
(2) \log \left(\frac{m_b}{m_c}\right)+\frac{23}{4} \log
\left(\frac{m_b}{m_c}\right)+\frac{\log
   ^2(2)}{2}\right.\nonumber\\
&&\left.+\frac{11 \log (2)}{2}+\frac{5 \pi ^2}{3}-19\right)\nonumber\\
&&+C_A\left(-\frac{1}{4} \log
^2\left(\frac{m_b}{m_c}\right)+\frac{3}{2} \log (2) \log
\left(\frac{m_b}{m_c}\right)+\frac{1}{2} \log
\left(\frac{m_b}{m_c}\right)+\frac{\log
   ^2(2)}{2}\right.\nonumber\\
&&\left.-5 \log (2)-\frac{\pi ^2}{8}+\frac{76}{9}\right)
\end{eqnarray}
\begin{eqnarray}
\frac{g^{NLO}(q^2)}{g^{LO}(q^2)}&=&1+\frac{\alpha_s}{4\pi}\{\frac{1}{3}
(11 C_A-2 \text{n}_f) \log
 (\frac{\mu^2}{2 \gamma\text{m}_c^2})-\frac{10\text{n}_f}{9}\nonumber\\
&&-\frac{C_A}{36 s-18}(9 s (2 s-1) \log ^2(s)+18 (2 s \log (2) (2
s-1)+1) \log (s)\nonumber\\
&&+3 \pi ^2 (s+2) (2 s-1)-2 s (-18 \log ^2(2) s+9 \log ^2(2)+45 \log
(2)+134)\nonumber\\
&&+9 (2 s-1)
   (\log (\gamma )-3) \log (\gamma )+18 s (2 s-1) (2
   \text{Li}_2(1-2 s)-\text{Li}_2(1-s))\nonumber\\
&&+63 \log
   (2)+134)\nonumber\\
&&\frac{C_F}{6 (1-2 s)^2 (s-1)}(6 (s^2-1) \log ^2(s) (1-2 s)^2+24
(s-1) s \text{Li}_2(1-2 s) (1-2 s)^2\nonumber\\
&&+3 (2 s (s (4 s (4 \log (2) s-8 \log (2)+3)+20 \log (2)-17)-4 \log
(2)+7)-1)
   \log (s)\nonumber\\
&&+(s-1) (6 \log (\gamma ) (\log (\gamma )-6 \log (2)+5) (1-2 s)^2+6
(2 s-9) \log ^2(2) (1-2 s)^2\nonumber\\
&&+(2 s-1) (-204 s+2 \pi ^2 (2
   s^2+s-1)+105)+6 (s (68 s-67)+16) \log (2))\nonumber\\
&&-12 (2 s^2-3 s+1)^2
   \text{Li}_2(1-s))\}
\end{eqnarray}
\begin{eqnarray}
\frac{g^{NLO}(0)}{g^{LO}(0)}&=&1+\frac{\alpha_s}{4\pi}\{\frac{1}{3}
(11 C_A-2 \text{n}_f) \log
 \left(\frac{2\mu^2}{m_bm_c}\right)-\frac{10\text{n}_f}{9}\nonumber\\
&&+C_F\left(\log ^2\left(\frac{m_b}{m_c}\right)-10 \log (2) \log \left(
\frac{m_b}{m_c}\right)+5 \log \left(\frac{m_b}{m_c}\right)
+9 \log ^2(2)\right.\nonumber\\
&&\left.+
7 \log (2)+\frac{\pi ^2}{3}-15\right)\nonumber\\
&&+C_A\left(-\frac{1}{2} \log ^2\left(\frac{m_b}{m_c}\right)+2 \log
(2) \log \left(\frac{m_b}{m_c}\right)+\frac{3}{2} \log
\left(\frac{m_b}{m_c}\right)-3 \log ^2(2)\right.\nonumber\\
&&\left.-\frac{3 \log
   (2)}{2}-\frac{\pi ^2}{3}+\frac{67}{9}\right)
\end{eqnarray}
\begin{eqnarray}
\frac{a_0^{NLO}(q^2)}{a_0^{LO}(q^2)}=\frac{a_+^{NLO}(q^2)}{a_+^{LO}(q^2)}=
\frac{g^{NLO}(q^2)}{g^{LO}(q^2)}
\end{eqnarray}
\begin{eqnarray}
\frac{a_0^{NLO}(0)}{a_0^{LO}(0)}=\frac{a_+^{NLO}(0)}{a_+^{LO}(0)}=
\frac{g^{NLO}(0)}{g^{LO}(0)}
\end{eqnarray}
\begin{eqnarray}
\frac{a_-^{NLO}(q^2)}{a_-^{LO}(q^2)}&=&1+\frac{\alpha_s}{4\pi}\{\frac{1}{3}
(11 C_A-2 \text{n}_f) \log
 (\frac{\mu^2}{2 \gamma\text{m}_c^2})-\frac{10\text{n}_f}{9}\nonumber\\
&&-\frac{C_A}{9 (s-1) (2 s-1)}(18 (2 \log (2) s^2-3 \log (2)
s+s+\log (2)) \log (s)\nonumber\\
&&+(s-1) (2 (-85+36 (-1+\log (2)) \log (2)) s+\pi ^2 (6 s-3)+18 (2
s-1) \log (2) \log (\gamma
   )\nonumber\\
&&-18 \log (2) (-1+2 \log (2))+85)+18 (s-1) (2 s-1) \text{Li}_2(1-2
s)\nonumber\\
&&-18 (s-1) (2 s-1)
   \text{Li}_2(1-s))\nonumber\\
&&\frac{C_F}{(2s^2-3s+1)^2}(\log ^2(s) (2 s^2-3 s+1)^2+4
\text{Li}_2(1-2 s) (2 s^2-3 s+1)^2\nonumber\\
&&-2 \text{Li}_2(1-s) (2 s^2-3 s+1)^2+(s (s (4 (3+4 \log
   (2)) s^2-8 (5+6 \log (2)) s\nonumber\\
&&+52 \log (2)+27)-4 (1+6 \log (2)))+4 \log (2)) \log
(s)+\frac{1}{3} (12 (-13\nonumber\\
&&+2 \log (2)+\log ^2(2))
   s^4-12 (-43+5 \log (2)+3 \log ^2(2)) s^3+(-597\nonumber\\
&&+54 \log (2)+26 \log (2)^2 ) s^2-3 (-95+6 \log ^2(2)+8\log
(2)) s\nonumber\\
&&-2 \pi ^2 (2
   s^2-3 s+1)^2-3 (2 s^2-3 s+1)^2 \log (\gamma )
   (\log (\gamma )+2 \log (2)-6)\nonumber\\
&&+3 \log ^2(2)+6 \log
   (2)-48))\end{eqnarray}
\begin{eqnarray}
\frac{a^{NLO}_-(0)}{a^{LO}_-(0)}&=&1+\frac{\alpha_s}{4\pi}\{\frac{1}{3}
(11 C_A-2 \text{n}_f) \log
 \left(\frac{2\mu^2}{m_bm_c}\right)-\frac{10\text{n}_f}{9}\nonumber\\
&&+C_F\left(-\log ^2\left(\frac{m_b}{m_c}\right)
+2 \log (2) \log \left(\frac{m_b}{m_c}\right)
+6 \log \left(\frac{m_b}{m_c}\right)+\log ^2(2)\right.\nonumber\\
&&\left.-6 \log (2)-\pi ^2-\frac{29}{2}\right)\nonumber\\
&&+C_A\left(-2 \log (2) \log \left(\frac{m_b}{m_c}\right)+6 \log
(2)-\frac{\pi ^2}{6}+\frac{67}{9}\right)
\end{eqnarray}


\newpage
\begin{thebibliography}{99}

\bibitem{exp1}T. Aaltonen et al. (CDF Collaboration), Phys. Rev.
Lett. {\bf100}, 182002 (2008); A. Abulencia et al. (CDF
Collaboration), Phy. Lett. {\bf97}, 012002 (2006); V. M. Abazoz et
al. (D0 Collaboration), Phys. Rev. Lett. {\bf102}, 092001 (2009).

\bibitem{bib1}I. Bediaga and J. H. Munoz, arXiv:1102.2190.

\bibitem{bib2}X. Liu, Z. J. Xiao, and C. D. Lu,
Phys. Rev. D {\bf81}, 014022 (2010).

\bibitem{Fac1}A. J. Buras, arXiv:hep-ph9806471.

\bibitem{Fac2}M. Beneke, G. Buchalla, M. Neubert and
C. T. Sachrajda, Phys. Rev. Lett. {\bf83}, 1914 (1999); Nucl. Phys.
{\bf B591}, 313 (2000); Nucl. Phys. {\bf B606}, 245 (2001).

\bibitem{wilson1}K. G. Wilson, Phys. Rev. {\bf179}, 1499 (1969); K. G.
Wilson and W. Zimmermann, Comm. Math. Phy. {\bf24}, 87 (1972).

\bibitem{wilson2}W. Zimmermann, in Proc. 1970 Brandeis
Summer Institute in Theor. Phys, (eds. S. Deser, M. Grisaru and H.
Pendleton), MIT Press, 1971, p.396; Ann. Phys. {\bf77}, 570 (1973).

\bibitem{CKM1}N. Cabibbo, Phys. Rev. Lett. {\bf10}, 531 (1963).

\bibitem{CKM2}M. Kobayashi and K. Maskawa, Pro. Theor. Phy. {\bf49}, 652
(1973).

\bibitem{NRQCD} G.T. Bodwin, E. Braaten, and
G. P. Lepage, Phys. Rev. D {\bf51}, 1125 (1995).

\bibitem{Bc1}G. Bell and Th. Feldmannb, Nuclear
Physics B (Proc. Suppl.) {\bf164}, 189 (2007).

\bibitem{Bc2} G. Bell, arXiv:0705.3133.

\bibitem{Chanowitz:1979zu}
  M.~S.~Chanowitz, M.~Furman and I.~Hinchliffe,
  Nucl.\ Phys.\ B {\bf 159}, 225 (1979).

\bibitem{'tHooft:1972fi}
  G.~'t Hooft and M.~J.~G.~Veltman,
  Nucl.\ Phys.\ B {\bf 44}, 189 (1972).

\bibitem{Petrelli:1997ge}
  A.~Petrelli, M.~Cacciari, M.~Greco, F.~Maltoni and M.~L.~Mangano,
  Nucl.\ Phys.\ B {\bf 514}, 245 (1998)

\bibitem{Charles:1998dr}
  J.~Charles, A.~Le Yaouanc, L.~Oliver, O.~Pene and J.~C.~Raynal,
  Phys.\ Rev.\  D {\bf 60}, 014001 (1999)


\bibitem{feynarts} T. Hahn, Comput. Phys. Commun. {\bf140}, 418 (2001).

\bibitem{feyncalc} R. Mertig, M. B\"ohm, and A. Denner, Comput.
 Phys. Commun. {\bf4}, 345 (1991).

\bibitem{looptools} T. Hahn and M. Perez-Victoria,
Comput. Phys. Commun. {\bf118}, 153 (1999).

\bibitem{Du:1988ws}
  D.~s.~Du and Z.~Wang,
  Phys.\ Rev.\  D {\bf 39}, 1342 (1989).

\bibitem{Colangelo:1992cx}
 P.~Colangelo, G.~Nardulli and N.~Paver,
  Z.\ Phys.\  C {\bf 57}, 43 (1993).

\bibitem{Kiselev:1993ea}
  V.~V.~Kiselev and A.~V.~Tkabladze,
  Phys.\ Rev.\  D {\bf 48}, 5208 (1993).

\bibitem{Kiselev:1999sc}
  V.~V.~Kiselev, A.~K.~Likhoded and A.~I.~Onishchenko,
  Nucl.\ Phys.\  B {\bf 569}, 473 (2000)
  Phys.\ Atom.\ Nucl.\  {\bf 63} (2000) 2123
  [Yad.\ Fiz.\  {\bf 63} (2000) 2219].

\bibitem{Nobes:2000pm}
  M.~A.~Nobes and R.~M.~Woloshyn,
  J.\ Phys.\ G {\bf 26}, 1079 (2000)

\bibitem{Ivanov:2000aj}
  M.~A.~Ivanov, J.~G.~Korner and P.~Santorelli,
  Phys.\ Rev.\  D {\bf 63}, 074010 (2001)

\bibitem{Kiselev:2002vz}
  V.~V.~Kiselev, A.~E.~Kovalsky and A.~K.~Likhoded,
  Nucl.\ Phys.\  B {\bf 585}, 353 (2000)
  V.~V.~Kiselev,
  arXiv:hep-ph/0211021

\bibitem{Ebert:2003cn}
  D.~Ebert, R.~N.~Faustov and V.~O.~Galkin,
  Phys.\ Rev.\  D {\bf 68}, 094020 (2003)
  D.~Ebert, R.~N.~Faustov and V.~O.~Galkin,
  Eur.\ Phys.\ J.\  C {\bf 32}, 29 (2003).
\bibitem{Ivanov:2005fd}
  M.~A.~Ivanov, J.~G.~Korner and P.~Santorelli,
  Phys.\ Rev.\  D {\bf 71}, 094006 (2005)
  [Erratum-ibid.\  D {\bf 75}, 019901 (2007)]

\bibitem{Hernandez:2006gt}
  E.~Hernandez, J.~Nieves and J.~M.~Verde-Velasco,
  Phys.\ Rev.\  D {\bf 74}, 074008 (2006).

\bibitem{Huang:2007kb}
  F.~Zuo and T.~Huang,
  Chin.\ Phys.\ Lett.\  {\bf 24}, 61 (2007)
  Eur.\ Phys.\ J.\  C {\bf 51}, 833 (2007).

\bibitem{Sun:2008ew}
  J.~F.~Sun, D.~S.~Du and Y.~L.~Yang,
  arXiv:0808.3619.

\bibitem{DSV}
  R.~Dhir, N.~Sharma and R.~C.~Verma,
  J.\ Phys.\ G {\bf 35}, 085002 (2008);
  R.~C.~Verma and A.~Sharma,
  Phys.\ Rev.\  D {\bf 65}, 114007 (2002);
  R.~Dhir and R.~C.~Verma,
  arXiv:0810.4284
\bibitem{Wang:2008xt}
  W.~Wang, Y.~L.~Shen, C.~D.~Lu,
  Phys.\ Rev.\  {\bf D79}, 054012 (2009).


\end{thebibliography}
\end{document}